\documentclass[journal]{IEEEtran}
\usepackage{cite}
\usepackage[pdftex]{graphicx}
\usepackage{epstopdf}
\usepackage{amsmath}
\usepackage{cases}
\usepackage[caption=false,font=footnotesize]{subfig}
\usepackage{fixltx2e}
\usepackage{amssymb}
\usepackage{mathtools}
\usepackage{bm}
\usepackage{multirow,booktabs}
\usepackage{color}
\usepackage[ruled]{algorithm2e}
\usepackage{mathrsfs}
\usepackage{amsthm,amsmath,amssymb}
\usepackage{threeparttable}
\usepackage{bbding}
\usepackage{wasysym}
\usepackage{nomencl}
\usepackage{etoolbox}
\usepackage{textcomp,mathcomp}
\usepackage[hyphens]{url}
\usepackage{breakurl}
\usepackage{arydshln}
\newcommand{\tabincell}[2]{\renewcommand\arraystretch{0.9}\begin{tabular}{@{}#1@{}}#2\end{tabular}}

\begin{document}

\title{Multi-Stage Expansion Planning for Decarbonizing Thermal Generation Supported Renewable Power Systems Using Hydrogen and Ammonia Storage}




\author{Zhipeng~Yu,~\IEEEmembership{Student Member,~IEEE},
        Jin~Lin,~\IEEEmembership{Member,~IEEE},
        Feng~Liu,~\IEEEmembership{Senior Member,~IEEE},
        Jiarong~Li,~\IEEEmembership{Member,~IEEE},
        Yingtian~Chi,
        Yonghua~Song,~\IEEEmembership{Fellow,~IEEE}, and
        Zhengwei~Ren
\thanks{This work was supported by the National Key R\&D Program of China (2021YFB4000500), National Natural Science Foundation of China (52207116, 52177092 and U22A20220) and China Postdoctoral Science Foundation (2022M711758). \emph{(Corresponding author: Jin Lin)}
}
\thanks{Z. Yu, J. Lin, F.Liu, J. Li, and Y. Chi are with the State Key Laboratory of Control and Simulation of Power Systems and Generation Equipment, Department of Electrical Engineering, Tsinghua University, Beijing 100087, China. And Jin Lin is also with Sichuan Energy Internet Research Institute, Tsinghua University, Chengdu, 610213, China. (e-mail: linjin@tsinghua.edu.cn) 
}
\thanks{Y. Song is with the Department of Electrical and Computer Engineering, University of Macau, Macau 999078, China, and also with the Department of Electrical Engineering, Tsinghua University, Beijing 100087, China. 
}
}
\maketitle

\begin{abstract}
  Large-scale centralized development of wind and solar energy and \emph{peer-to-grid} transmission of renewable energy source (RES) via high voltage direct current (HVDC) has been regarded as one of the most promising ways to achieve goals of peak carbon and carbon neutrality in China. Traditionally, large-scale thermal generation is needed to economically support the load demand of HVDC with a given profile, which in turn raises concerns about carbon emissions. To address the issues above, hydrogen energy storage system (HESS) and ammonia energy storage system (AESS) are introduced to gradually replace thermal generation, which is represented as a multi-stage expansion planning (MSEP) problem. Specifically, first, HESS and AESS are established in the MSEP model with carbon emission reduction constraints, and yearly data with hourly time resolution are utilized for each stage to well describe the intermittence of RES. Then, a combined Dantzig-Wolfe decomposition (DWD) and column generation (CG) solution approach is proposed to efficiently solve the large-scale MSEP model. Finally, a real-life system in China is studied. The results indicate that HESS and AESS have the potential to handle the intermittence of RES, as well as the monthly imbalance between RES and load demand. Especially under the goal of carbon neutrality, the contribution of HESS and AESS in reducing levelized cost of energy (LCOE) reaches 12.28\% and 14.59\%, respectively, which finally leads to a LCOE of 0.4324 RMB/kWh.
\end{abstract}

\begin{IEEEkeywords}
power system decarbonization, multi-stage expansion planning, hydrogen and ammonia, long duration energy storage, levelized cost of storage
\end{IEEEkeywords}

\section{Introduction}
\label{sec:intro}

\subsection{Background and Motivation}
\IEEEPARstart{T}{he} pressure of $\mathrm{CO_2}$ emission reduction in the energy sector is considerable, especially since China has established a national goal of peaking $\mathrm{CO_2}$ emissions before 2030, as well as carbon neutrality by 2060 \cite{li2021quantitative,China-carbon-neutrality-web}. Large-scale centralized development of renewable energy source (RES) and \emph{peer-to-grid} transmission of RES via high voltage direct current (HVDC) has been identified as one of the most significant ways to achieve peak carbon and carbon neutrality goals in China \cite{liu2012economic,li2018recent,li2020engineering,li2021make}.

However, thermal generation is considered for the economics of power delivery. Currently, thermal generation usually accounts for 50\% of load demand in planned UHV projects, which follows the rule in \cite{CarbonPeak2030} that \emph{the proportion of renewable energy in new UHV channels shall not be less than 50\% in principle}. According to 14th five-year renewable energy development plan \cite{145RED2022}, nine ultra-high voltage direct current (UHVDC) projects ($\pm $ 800kV, 8GW), such as Jinshang-Hubei, Longdong-Shandong, and Hami-Chongqing, and three ultra-high voltage alternating current (UHVAC) projects (1000kV), such as Datong-Tianjing and Sichuan-Chongqing, are planned, which is known as the layout of \emph{3-UHVAC\&9-UHVDC}. Assuming that the full load hours (FLH) of UHV is 5000 hours, of which thermal power accounts for 50\%, then the annual thermal power generation of the above-mentioned UHV projects reaches 240 billion kWh, and the corresponding annual carbon emission is 200 Mt. This, on the contrary, raises new concerns about carbon emissions in power systems.

Constrained by the carbon emission reduction process, percentage of thermal plants in the overall electricity generation should be gradually reduced. One effective way is to apply long duration energy storage (LDES) to achieve stable generation of RES, and seasonal balance between RES and load demand \cite{jiang2021modeling,zhuo2022cost,zhang2022efficient,jiang2022renewable}. Specifically, HESS and AESS are feasible LDES technologies with seasonal regulation abilities \cite{song2023utilization,li2019optimal}. The above process can be described as a multi-stage expansion planning (MSEP) model for decarbonizing thermal generation supported renewable power systems. This is the main motivation and research work of this paper.



\subsection{Literature Review}
Existing works \cite{zhang2022efficient,jiang2022renewable,carvallo2023multi,zhou2023multiple,li2022coordinated,huang2022nested} have studied the capacity expansion planning of renewable power systems. \cite{zhang2022efficient} proposes a planning model considering a year-round hourly operation, i.e., the monthly fluctuation characteristics of RES are preserved as much as possible by generating several typical days clustered in each month. \cite{jiang2022renewable} further quantify the seasonal imbalance risk by introducing the conditional value at risk (CVaR) method. \cite{carvallo2023multi} studies decarbonization and decontamination planning for isolated systems; and wet, medium, and dry scenarios for hydro scenarios and three scenarios of high, medium, and low wind resource availability are generated, respectively, and representative days (RD) are finally derived by combining hydro and wind scenarios. There are other RD generation methods, such as four typical days representing spring, summer, autumn, and winter in \cite{zhou2023multiple,li2022coordinated}. In addition, \cite{huang2022nested} further distinguishes the typical day between weekdays and weekends.

The above research works based on RD focus on addressing the stochasticity and volatility of RES. However, ignoring the description of RES' intermittence leads to planning results that tend to allocate little or no LDES. Similarly, \cite{wu2022planning} proposes a planning model for a hydrogen-based carbon-free system by using yearly data. The case study indicates that the results based on yearly data can fully consider the energy complementary between days or even seasons, leading to a much lower cost than in planning results obtained by RD.

\cite{zheng2022incorporating,yu2023optimal,yu2023optimal_Iso} also utilize the yearly data in operation model for capacity optimization. Hydrogen production and storage are established in \cite{zheng2022incorporating}--\cite{yu2023optimal}, while hydrogen-to-power by fuel cell (FC) is further introduced in \cite{yu2023optimal_Iso}, which makes it strictly an HESS. Additionally, \cite{haggi2022proactive}--\cite{wen2023data} identify the roles of HESS for daily load shifting by operation simulation. However, why HESS and AESS should be allocated is still unclear.

To the best of our knowledge, there is little or no work on multi-stage expansion planning (MSEP) considering both decarbonization and LDES. Although existing RD-based methods can reduce the scale to efficiently solve the optimization problem, the intermittence of RES and seasonal imbalance are not well considered leading to little demand for LDES. However, if yearly data with hourly time resolution are directly utilized at each planning stage, the MSEP model will exhibit large-scale optimization problems with millions or even tens of millions of variables and constraints. How to efficiently solve such large-scale problems remains underexplored.


\subsection{Contributions}
\label{sec:contributions}
To fill the above-mentioned gaps, first, we propose an MSEP model considering the carbon emission reduction constraints. And HESS and AESS are established as LDES. Second, a combined Dantzig-Wolfe decomposition (DWD) and column generation (CG) method is proposed to efficiently solve the large-scale MSEP model. Then, the LCOS assessment method for BESS, HESS, and AESS based on virtual internal trading is introduced. Finally, a real-life UHVDC of Longdong--Shandong in China is studied.

Specifically, the following contributions are made in this paper:

1) {\bf{Modeling Level}}: A multi-stage expansion planning model for decarbonizing the thermal generation supported renewable power system is established for the first time, with HESS and AESS gradually replacing thermal generation. It also provides a novel and feasible path for economical and low-carbon large-scale renewable energy transmission.

2) {\bf{Solution Approach Level}}: A combined Dantzig-Wolfe decomposition (DWD) and column generation (CG) solution approach is proposed to accurately and efficiently solve the large-scale problem with acceptable and controllable computational burden. Compared to existing RD-based methods, the proposed DWD-CG directly solves the problem formulated using yearly data for each stage, accurately describing the demand for LDES.  

3) {\bf{Application Level}}: Using data from real-life system, the planning results indicate that HESS and AESS differ in regulation time scales, i.e., HESS has the ability and characteristics of daily and weekly regulation, while HESS focuses on monthly and seasonal regulation. It is also revealed that HESS and AESS have the potential to handle the intermittence of RES and the seasonal imbalance between RES and load demand. It is better to configure both HESS and AESS than to configure HESS separately. Specifically, under the goal of carbon neutrality, the contribution of HESS and AESS in reducing LCOE reaches $12.28\%$ and $14.59\%$, respectively.

\section{Problem Description}
\label{sec:problem_description}
In this section, first, the corresponding configuration of the proposed thermal generation supported renewable power system with HESS and AESS is presented. Then, the reason of using HESS and AESS to replace thermal generation is illustrated.

\subsection{Configuration of the Proposed Thermal Generation Supported Renewable Power System with HESS and AESS}
\begin{figure}[t]
  \centering
  \includegraphics[width=3.46in]{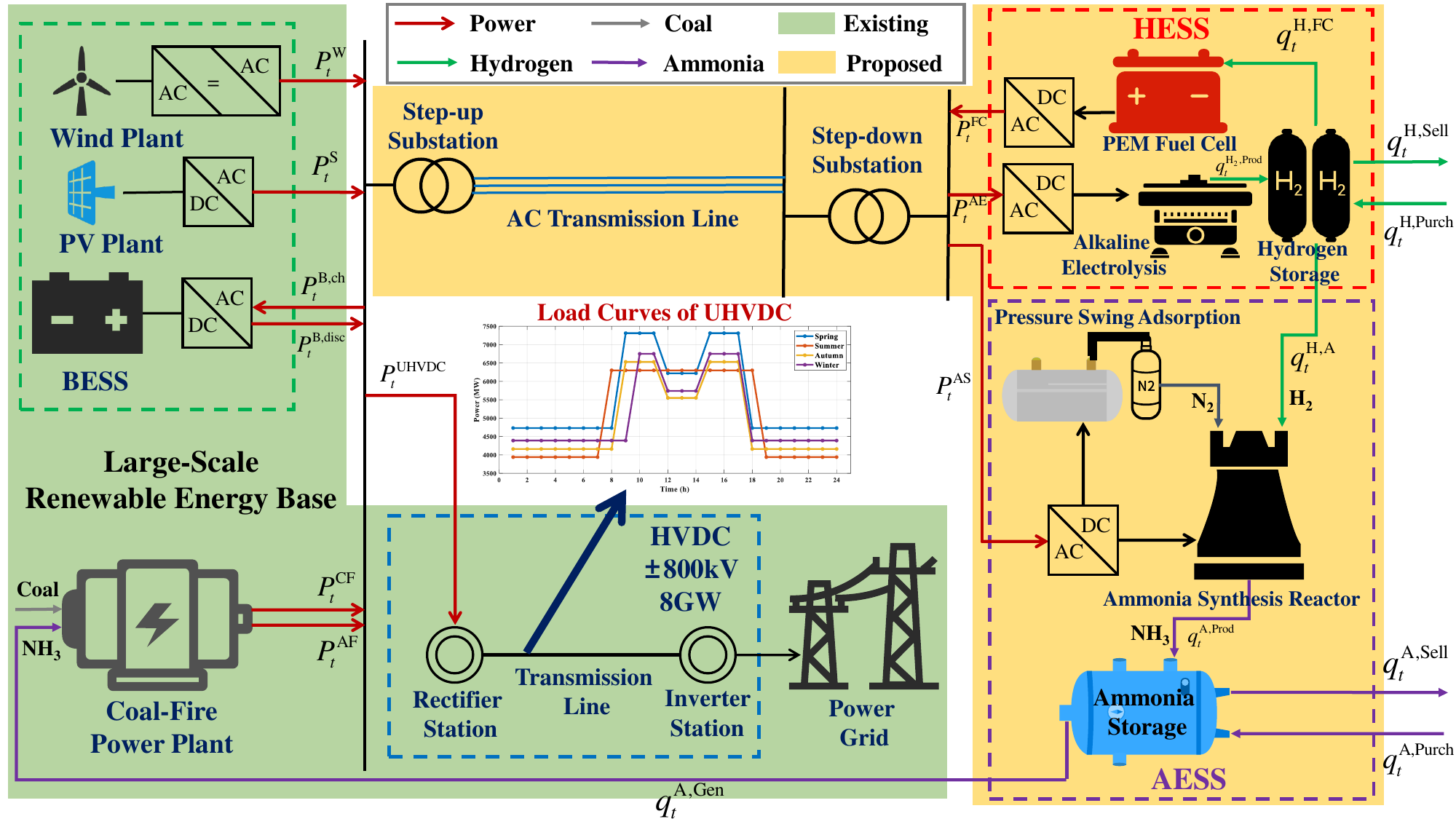}
  \caption{Topology of thermal generation supported renewable power systems with HESS and AESS.}
  \label{fig:System_Topology}
\end{figure}

Existing thermal generation supported renewable power systems are presented as the green region shown in Fig. \ref{fig:System_Topology}, including wind turbine (WT), photovoltaic (PV), thermal generation by coal-fired power plant (CFPP), battery energy storage systems (BESS), and UHVDC.

Furthermore, HESS and AESS are introduced, as shown in the orange region in Fig. \ref{fig:System_Topology}. Specifically, HESS consists of three parts: \emph{hydrogen production} using alkaline electrolysis (AE) technology; \emph{hydrogen storage} using a buffer tank; and \emph{hydrogen generation} using proton exchange membrane fuel cell (PEMFC). And AESS consists of three parts: \emph{ammonia production}, including nitrogen produced by pressure swing adsorption (PSA) and ammonia synthesis with Haber Bosch synthesis (HBS); \emph{ammonia storage} using buffer tank; and \emph{ammonia generation} using coal-fired thermal units.

\subsection{The Feasibility and Advantage of Replacing Thermal Generation with HESS and HESS}

\subsubsection{Lower Capital Expenditure (CAPEX) of Hydrogen and Ammonia Storage by Buffer Tank Compared to Electricity Storage by Battery }
Based on the investment parameter \cite{yu2023optimal} of electricity storage by Li-battery, hydrogen storage in high pressure gas cylinders, and ammonia storage in low-temperature fluid containers, we find that CAPEX (converting to unit energy) of electricity, hydrogen, and ammonia storages are 500 RMB/MJ, 23.2 RMB/MJ, and 0.18 RMB/MJ, respectively. It means that the investment cost of one minute level electricity storage is equal to that of twenty minutes level hydrogen storage and two days level ammonia storage. Therefore, HESS and AESS are better suited to handle long-term and low-frequency regulation than BESS since the CAPEX of hydrogen and ammonia storage is several orders of magnitude lower than electricity storage.

\subsubsection{Higher Utilization Rate of Thermal Generation Units by Ammonia-Fired Generation}
Since ammonia-fired generation can use existing thermal generation units, it can delay the retirement of thermal generation units and improve the utilization rate of facilities. Therefore, on the one hand, the CAPEX of AESS is further reduced, which makes AESS more economically feasible; on the other hand, the utilization rate of thermal generation units is improved, which makes GenCos willing to invest in thermal generation at present. This is conducive to realizing large-scale RES transmission by HVDC from the west to the east of China. 

In addition, under the tightened carbon emission reduction constraints, HESS and AESS will gradually replace thermal generation, which exhibits a multi-stage expansion and decarbonization planning model in Section \ref{sec:MSEDP}. 

\section{Multi-Stage Expansion and Decarbonization Planning Model}
\label{sec:MSEDP}
In this section, first, the MSEP model is introduced. Then, calculation methods for performance indices are presented. 

\subsection{Mathematical Model of MSEP}
\label{sec:MSEP}
\subsubsection{Objective Function}
The objective function of the MSEP model, represented in (\ref{eq:NPC}), is the net present cost (NPC) of the whole system, which consists of three parts: 

\emph{a) The present value of cost (PVC)}, including initial investment of all facilities $PVC_s^{\rm{inv}}$ in (\ref{eq:NPC_inv}), fixed operation and maintenance (O\&M) cost $PVC_s^{\rm{O\&M}}$ in (\ref{eq:NPC_OM}), retirement cost of CFPP $PVC_{s}^{\rm{reti}}$ in (\ref{eq:NPC_reti}), cost of buying coal for CFPP $PVC_{s}^{\rm{coal}}$ in (\ref{eq:NPC_coal}), cost of degeneration for BESS $PVC_{s}^{\rm{deg}}$ in (\ref{eq:NPC_deg}) \cite{zhao2020stochastic}, cost of buying green hydrogen $PVC_{s}^{\rm{H,purch}}$ in (\ref{eq:NPC_hydrogen}), and cost of buying green ammonia $PVC_{s}^{\rm{A,purch}}$ in (\ref{eq:NPC_ammonia}). 

\emph{b) The present value of revenue (PVR)}, including revenue from selling hydrogen $PVR_{s}^{\rm{H,sell}}$ in (\ref{eq:NPR_hydrogen}) and revenue from selling ammonia $PVR_{s}^{\rm{A,sell}}$ in (\ref{eq:NPR_ammonia}). 

\emph{c) The present value of savings (PVS)}, including the savings of all facilities at the end of the planning horizon, denoted as $PVS_{s}$ in (\ref{eq:NPC_Savings}). 

Note that present value is related to the interest rate $r = 8\%$, calculated by $\delta _1(y)$, $\delta _2(y)$, and $\delta _3(y)$ in (\ref{eq:Set_Coeff}). In addition, coefficients of facilities' savings $I_s^{j,\rm{sav}}$, the set of planning stages $\mathbb{S}$, the set of time indices $\mathbb{T}$, and the set of facilities $\Omega_{\rm{F}}$ are also presented in (\ref{eq:Set_Coeff}).
\begin{subequations}
  \begin{align}
    &NPC = \sum \limits_{s \in \mathbb{S}} PVC_{s}^{\rm{inv}} + PVC_{s}^{\rm{O\&M}}+ PVC_{s}^{\rm{reti}} + PVC_{s}^{\rm{coal}}  \nonumber \\ 
    & \qquad \qquad  + PVC_s^{\rm{deg}} + PVC_{s}^{\rm{H,purch}} + PVC_{s}^{\rm{A,purch}} \nonumber \\ 
    & \qquad \qquad - PVR_{s}^{\rm{H,sell}}- PVR_{s}^{\rm{A,sell}} -PVS_{s}\label{eq:NPC} \\
    &PVC_s^{\rm{inv}} = \delta_{1}((s-1)\Delta S+1)\sum_{j \in {\Omega}_{\rm{F}}}  I_s^{j,\rm{init}}  \Delta C_s^j  \label{eq:NPC_inv}\\
    &PVC_s^{\rm{O\&M}} = \delta_{2}(s) \sum_{j \in {\Omega}_{\rm{F}}} I_s^{j,\rm{O\&M}} I_s^{j,\rm{init}}\Delta C_s^j \label{eq:NPC_OM} \\
    &PVC_s^{\rm{reti}} = \delta_{1}((s-1)\Delta S+1) I_s^{\rm{CFPP,reti}} \Delta C_s^{\rm{CFPP,reti}} \label{eq:NPC_reti} \\
    &PVC_{s}^{\rm{coal}} = \delta_{3}(s) \pi_{s,t}^{\rm{coal}} \Delta T \sum_{t \in \mathbb{T}} \gamma_{\rm{P2C}}P_{s,t}^{\rm{CF}}  \label{eq:NPC_coal} \\
    &PVC_s^{\rm{deg}} = \lambda_{\rm{deg}} \Delta T \sum_{t \in \mathbb{T}} P_{s,t}^{\rm{B,disc}}  \label{eq:NPC_deg} \\
    &PVC_{s}^{\rm{H,purch}} = \delta_{3}(s) \pi_{s,t}^{\rm{H,purch}} \Delta T \sum_{t \in \mathbb{T}} q_{s,t}^{\rm{H,purch}}  \label{eq:NPC_hydrogen} \\
    &PVC_{s}^{\rm{A,purch}} = \delta_{3}(s) \pi_{s,t}^{\rm{A,purch}} \Delta T \sum_{t \in \mathbb{T}} q_{s,t}^{\rm{A,purch}}  \label{eq:NPC_ammonia} \\
    &PVR_{s}^{\rm{H,sell}} = \delta_{3}(s) \pi_{s,t}^{\rm{H,sell}} \Delta T \sum_{t \in \mathbb{T}} q_{s,t}^{\rm{H,sell}}  \label{eq:NPR_hydrogen} \\
    &PVR_{s}^{\rm{A,sell}} = \delta_{3}(s) \pi _{s,t}^{\rm{A,sell}} \Delta T \sum_{t \in \mathbb{T}} q_{s,t}^{\rm{A,sell}}  \label{eq:NPR_ammonia} \\
    &PVS_s = \delta_{1}(S \Delta S) \sum_{j \in {\Omega}_{\rm{F}}} I_s^{j,\rm{sav}} I_s^{j,\rm{init}}\Delta C_s^j \label{eq:NPC_Savings}  \\
    &\delta_{1}(y) = (1+r)^{-y} , \; \delta_{2}(s) = \sum\nolimits_{y=(s-1)\Delta S+1}^{S \Delta S} \delta_{1}(y),  \nonumber\\
    &\delta_{3}(s) = \sum\nolimits_{y=(s-1)\Delta S+1}^{s \Delta S} \delta_{1}(y), \nonumber \\
    &I_s^{j,\rm{sav}} = \max{\left\{1-\frac{(S-s+1)\Delta S-1}{LT^j-1},0\right\}}, \nonumber \\
    &\mathbb{S} = \left\{1,2,\ldots,S \right \},\ \mathbb{T} = \left\{0,1,\ldots,N-1  \right\} \nonumber \\
    &\Omega_{\rm{F}} = \left\{ \rm{W},\rm{S},\rm{CFPP},\rm{B},\rm{AE},\rm{HS},\rm{FC},\rm{ASyn},\rm{ASto}\right\}   \label{eq:Set_Coeff} 
  \end{align}
\end{subequations}

\subsubsection{Constraints of Facilities' Capacities}
The coupling of different stages is reflected in (\ref{eq:Capa_1}), i.e., the capacity in the current stage $C_s^j$ is determined by the capacity in the previous stage $C_{s-1}^j$ and the increased capacity at current stage $\Delta C_s^j$. And for CFPP, the retired capacity in the current stage $\Delta C_s^{\rm{CFPP,reti}}$ needs additional consideration. Furthermore, upper and lower limits of capacities are presented in (\ref{eq:Capa_2})--(\ref{eq:Capa_3}).
\begin{subequations}
  \begin{align}
    &C_s^j = C_{s-1}^j + \Delta C_s^j , \ \forall j \in \Omega_{\rm{F}} \backslash \left\{\rm{CFPP}\right\}, \nonumber\\
    &C_s^{\rm{CFPP}} = C_{s-1}^{\rm{CFPP}} + \Delta C_s^{\rm{CFPP}} - \Delta C_s^{\rm{CFPP,reti}} , \forall s \in \mathbb{S}\label{eq:Capa_1} \\
    &0 \leq C_s^j \leq \overline{C_s^j}, \ \forall j \in \Omega_{\rm{F}}, \forall s \in \mathbb{S} \label{eq:Capa_2} \\ 
    &\Delta C_s^j \geq 0,\ \Delta C_s^{\rm{CFPP,reti}} \geq 0, \ \forall j \in \Omega_{\rm{F}}, \forall s \in \mathbb{S} \label{eq:Capa_3}
  \end{align}
\end{subequations}

\subsubsection{Constraints of HESS}
The operation model of HESS is presented in (\ref{eq:D_AE_1})--(\ref{eq:D_FC_2}), including converting electrical power $P_{s,t}^{\rm{AE}}$ to hydrogen $q_{s,t}^{\rm{H,A}}$ in (\ref{eq:D_AE_1}), variation range limits of AE in (\ref{eq:D_AE_2}), the state space equation of HS in (\ref{eq:D_HS_1}), the state of charge (SOC) constraints of HS in (\ref{eq:D_HS_2})--(\ref{eq:D_HS_3}), converting hydrogen $q_{s,t}^{\rm{H,FC}}$ to power $P_{s,t}^{\rm{FC}}$ in (\ref{eq:D_FC_1}), and variation range limits of FC in (\ref{eq:D_FC_2}).
\begin{subequations}
  \begin{align}
    &P_{s,t}^{\rm{AE}} = \kappa_{\rm{AE}} q_{s,t}^{\rm{H,prod}}   \label{eq:D_AE_1}  \\
    &\underline{\eta}^{\rm{AE}} C_s^{\rm{AE}} \leq P_{s,t}^{\rm{AE}} \leq \overline{\eta}^{\rm{AE}} C_s^{\rm{AE}} \label{eq:D_AE_2}\\
    &n_{s,t+1}^{\rm{HS}} = n_{s,t}^{\rm{HS}} +  q_{s,t}^{\rm{H,prod}} \Delta T + q_{s,t}^{\rm{H,pruch}}\Delta T  \nonumber\\ 
    &\qquad \quad \; \ - q_{s,t}^{\rm{H,FC}} \Delta T -  q_{s,t}^{\rm{H,sell}} \Delta T - q_{s,t}^{\rm{H,A}} \Delta T \label{eq:D_HS_1} \\
    &\underline{\eta}^{\rm{HS}} C_s^{\rm{HS}} \leq n_{s,t}^{\rm{HS}} \leq \overline{\eta}^{\rm{HS}} C_s^{\rm{HS}} \label{eq:D_HS_2} \\
    &\left. n_{s,t}^{\rm{HS}} \right| _{t=0} = \left. n_{s,t}^{\rm{HS}} \right|_{t=N} \label{eq:D_HS_3}\\
    &P_{s,t}^{\rm{FC}} = \kappa_{\rm{FC}} q_{s,t}^{\rm{H,FC}}   \label{eq:D_FC_1} \\
    &\underline{\eta}^{\rm{FC}} C_s^{\rm{FC}} \leq P_{s,t}^{\rm{FC}} \leq \overline{\eta}^{\rm{FC}} C_s^{\rm{FC}},\forall s \in \mathbb{S}, \forall t \in \mathbb{T} \label{eq:D_FC_2}
  \end{align}
  \end{subequations}

\subsubsection{Constraints of AESS}
The operation model of AESS is presented in (\ref{eq:D_AS_1})--(\ref{eq:D_AS_10}). (\ref{eq:D_AS_1}) represents the process of converting electrical power $P_{s,t}^{\rm{AS}}$ to hydrogen $q_{s,t}^{\rm{H,A}}$. (\ref{eq:D_AS_2}) is the dynamic operation model of ammonia synthesis based on our previous works \cite{yu2023optimal,yu2023optimal_Iso}. The variation range and ramping limits of AS are shown in (\ref{eq:D_AS_3}) and (\ref{eq:D_AS_4}), respectively,  with the definition of the rated work condition of AS $q_{s}^{\rm{H,r}} $ in (\ref{eq:D_AS_5}). The relationship between hydrogen consumption and ammonia production by AS is presented in (\ref{eq:D_AS_6}).  (\ref{eq:D_AS_7}) is the state space equation of ASto with SOC constraints in (\ref{eq:D_AS_8})--(\ref{eq:D_AS_9}). Finally, (\ref{eq:D_AS_10}) represents the ammonia-fired process by converting ammonia $q_{s,t}^{\rm{A,gen}}$ to power $P_{s,t}^{\rm{AF}}$ using the existing coal-fired units.
\begin{subequations}
  \begin{align}
    &P_{s,t}^{\rm{AS}} = \kappa_{\rm{AS}} q_{s,t}^{\rm{H,A}}\label{eq:D_AS_1} \\
    &q_{s,\tau}^{\rm{H,A}} = q_{s,k}^{\rm{H,QSS}} + \left( q_{s,k}^{\rm{H,QSS}} - q_{s,k+1}^{\rm{H,QSS}} \right) e^{-\frac{\tau}{T_{\rm{trans}}}}, \nonumber \\
     &\forall k \in \mathbb{K}, \forall \tau \in \left[ k \Delta T_{\rm{AS}},(k+1) \Delta T_{\rm{AS}}\right) \label{eq:D_AS_2} \\
    &\underline{\eta}^{\rm{AS}} q_{s}^{\rm{H,r}} \leq q_{s,t}^{\rm{H,A}} \leq \overline{\eta}^{\rm{AS}} q_{s}^{\rm{H,r}} \label{eq:D_AS_3}\\
    &r_{-}^{\rm{AS}} q_{s}^{\rm{H,r}} \leq q_{s,t+1}^{\rm{H,A}} -q_{s,t}^{\rm{H,A}}  \leq r_{+}^{\rm{AS}} q_{s}^{\rm{H,r}} \label{eq:D_AS_4}\\
    &q_{s}^{\rm{H_2,r}} = C_s^{\rm{AS}}/(8000 \gamma_{\rm{H2A}}) \label{eq:D_AS_5} \\
    &q_{s,t}^{\rm{A,prod}} = \gamma_{\rm{H2A}} q_{s,t}^{\rm{H,A}} \label{eq:D_AS_6}\\ 
    &m_{s,t+1}^{\rm{ASto}} - m_{s,t}^{\rm{ASto}} =  q_{s,t}^{\rm{A,prod}} \Delta T + q_{s,t}^{\rm{A,pruch}}\Delta T  \nonumber\\ 
    &\qquad \qquad \qquad \quad  - q_{s,t}^{\rm{A,gen}} \Delta T -  q_{s,t}^{\rm{A,sell}} \Delta T \label{eq:D_AS_7} \\
    &\underline{\eta}^{\rm{ASto}} C_s^{\rm{ASto}} \leq n_{s,t}^{\rm{ASto}} \leq \overline{\eta}^{\rm{ASto}} C_s^{\rm{ASto}} \label{eq:D_AS_8} \\
    &\left. m_{s,t}^{\rm{ASto}} \right|_{t=0} = \left. m_{s,t}^{\rm{ASto}} \right|_{t=N} \label{eq:D_AS_9} \\
    &P_{s,t}^{\rm{AF}} = \gamma_{\rm{A2P}}q_{s,t}^{\rm{A,gen}},\forall s \in \mathbb{S}, \forall t \in \mathbb{T}  \label{eq:D_AS_10} 
  \end{align}
  \end{subequations}

\subsubsection{Constraints of Hydrogen and Ammonia Trading}
The external trading model of hydrogen and ammonia is listed in (\ref{eq:HA_trading_1})--(\ref{eq:HA_trading_4}), which follows the principle of an annual contract with a fixed trading mode at every stage. 
\begin{subequations}
  \begin{align}
    &q_{s,t+1}^{\rm{H,purch}} - q_{s,t}^{\rm{H,purch}} = 0 \label{eq:HA_trading_1} \\
    &q_{s,t+1}^{\rm{H,sell}} - q_{s,t}^{\rm{H,sell}} = 0 \label{eq:HA_trading_2} \\
    &q_{s,t+1}^{\rm{A,purch}} - q_{s,t}^{\rm{A,purch}} = 0 \label{eq:HA_trading_3} \\
    &q_{s,t+1}^{\rm{A,purch}} - q_{s,t}^{\rm{A,purch}} = 0,\ \forall s \in \mathbb{S}, \  \forall t \in \mathbb{T} \qquad  \label{eq:HA_trading_4}
  \end{align}
  \end{subequations}

\subsubsection{Constraints of BESS}
The operation model of BESS is presented in (\ref{eq:D_BES_1})--(\ref{eq:D_BES_4}), including the state space equation of BESS in (\ref{eq:D_BES_1}), SOC constraints of BESS in (\ref{eq:D_BES_2})--(\ref{eq:D_BES_3}), and charging/discharging power constraints in (\ref{eq:D_BES_4}).
\begin{subequations}
  \begin{align}
    &E_{s,t+1}^{\rm{B}} = (1-\xi_{\rm{B}})E_{s,t}^{\rm{B}} + (\eta_{\rm{B}}P_{s,t}^{\rm{B,ch}} - \frac{1}{\eta_{\rm{B}}} P_{s,t}^{\rm{B,disc}} ) \Delta T   \label{eq:D_BES_1} \\
    &\underline{\eta}^{\rm{B}} C_s^{\rm{B}} \leq E_{t}^{\rm{B}} \leq \overline{\eta}^{\rm{B}} C_s^{\rm{B}} \label{eq:D_BES_2} \\
    &\left. E_{s,t}^{\rm{B}} \right|_{t=0} = \left. E_{s,t}^{\rm{B}} \right|_{t=N} \label{eq:D_BES_3} \\
    &0 \leq P_{s,t}^{\rm{B,ch}},P_{s,t}^{\rm{B,disc}} \leq \frac{C_s^{\rm{B}}}{H_{\rm{B}}} \label{eq:D_BES_4} 
  \end{align}
  \end{subequations}

\subsubsection{Constraints of CFPP}
The operation model of CFPP is presented in (\ref{eq:CFPP_1})--(\ref{eq:CFPP_2}). (\ref{eq:CFPP_1}) represents that power of CFPP $P_{s,t}^{\rm{CFPP}}$ consists of coal-fired power $P_{s,t}^{\rm{CF}}$ and ammonia-fired power $P_{s,t}^{\rm{AF}}$. The variation range limits of CFPP are shown in (\ref{eq:CFPP_2}).
\begin{subequations}
  \begin{align}
    &P_{s,t}^{\rm{CFPP}} = P_{s,t}^{\rm{CF}} + P_{s,t}^{\rm{AF}} \label{eq:CFPP_1}\\
    &\underline{\eta}^{\rm{CFPP}} C_{s}^{\rm{CFPP}} \leq P_{s,t}^{\rm{CFPP}}  \overline{\eta}^{\rm{CFPP}} C_{s}^{\rm{CFPP}}, \forall s \in \mathbb{S},  \forall t \in \mathbb{T}  \label{eq:CFPP_2}
  \end{align}
  \end{subequations}

\subsubsection{Constraints of System Integration}
The system integration constraints are presented in (\ref{eq:D_RG})--(\ref{eq:pcurt}). (\ref{eq:D_RG}) reflects the relationship between renewable power generation and installed capacities of renewable generators. (\ref{eq:D_Power_Balance}) represents the hourly power balance of the system. The curtailment of renewable power generation $P_{s,t}^{\rm{curt}}$ should be nonnegative in (\ref{eq:pcurt}).
\begin{subequations}
  \begin{align}
    &P_{s,t}^{j} = C_s^{j} P_{t}^{j,\rm{sta}} , j \in \left\{\rm{W},\rm{S} \right\}   \label{eq:D_RG}\\
    &P_{s,t}^{\rm{W}} + P_{s,t}^{\rm{S}} + P_{s,t}^{\rm{CF}} + P_{s,t}^{\rm{B,disc}} + P_{s,t}^{\rm{FC}} + P_{s,t}^{\rm{AF}} \nonumber \\
    &= P_{s,t}^{\rm{UHVDC}} + P_{s,t}^{\rm{AE}} +P_{s,t}^{\rm{AS}}  + P_{s,t}^{\rm{B,ch}} + P_{s,t}^{\rm{curt}}  \label{eq:D_Power_Balance} \\
    &P_{s,t}^{\rm{curt}} \geq 0, \forall s \in \mathbb{S}, \forall t \in \mathbb{T} \label{eq:pcurt} 
  \end{align}
\end{subequations}

\subsubsection{Constraints of Carbon Emission Reduction}
Carbon emission reduction targets should be met at every planning stage, denoted as (\ref{eq:CER_1}). Initial carbon emissions $CE_{0}$ are calculated by (\ref{eq:CER_2}), which follows the rule in \cite{CarbonPeak2030} that \emph{the proportion of renewable energy in new UHVDC channels shall not be less than 50\% in principle}.
\begin{subequations}
\begin{align}
  &\mu_{\rm{CF}} \sum \limits_{t \in \mathbb{T}} P_{s,t}^{\rm{CF}} \leq CE_{0} \left( 1-r_{s}^{\rm{CER}}\right), \forall s \in \mathbb{S} \label{eq:CER_1}\\
  &CE_{0} = 50\%\mu_{\rm{CF}} \sum \limits_{t \in \mathbb{T}} P_{1,t}^{\rm{UHVDC}} \label{eq:CER_2}
\end{align}
\end{subequations}

\subsubsection{Decision Variables}
Finally, decision variables are presented in (\ref{eq:D_dv_1})--(\ref{eq:D_dv_4}), including capacity-related variables in (\ref{eq:D_dv_1}), electrical power related operation variables in (\ref{eq:D_dv_2}), hydrogen and ammonia flow related operation variables in (\ref{eq:D_dv_3}), and SOC-related state variables in (\ref{eq:D_dv_4}).
\begin{subequations}
\begin{align}
  &C_s^{j},\Delta C_s^{j}, \Delta C_s^{\rm{CFPP,reti}},\forall j \in {\Omega}_{\rm{F}},\forall s \in \mathbb{S} \label{eq:D_dv_1} \\
  &P_{s,t}^{\rm{W}}, P_{s,t}^{\rm{S}}, P_{s,t}^{\rm{CFPP}}, P_{s,t}^{\rm{UHVDC}}, P_{s,t}^{\rm{AE}}, P_{s,t}^{\rm{AS}}, P_{s,t}^{\rm{FC}} , P_{s,t}^{\rm{B,disc}}, \nonumber \\
  &P_{s,t}^{\rm{B,ch}},P_{s,t}^{\rm{curt}},P_{s,t}^{\rm{CF}},P_{s,t}^{\rm{AF}}, \forall s \in \mathbb{S}, \forall t \in \mathbb{T}  \label{eq:D_dv_2} \\
  &q_{s,t}^{\rm{H,prod}}, q_{s,t}^{\rm{H,purch}}, q_{s,t}^{\rm{H,FC}}, q_{s,t}^{\rm{H,sell}}, q_{s,t}^{\rm{H,A}} \nonumber \\
  &q_{s,t}^{\rm{A,prod}}, q_{s,t}^{\rm{A,purch}}, q_{s,t}^{\rm{A,gen}}, q_{s,t}^{\rm{A,sell}}, \forall s \in \mathbb{S},\forall t \in \mathbb{T}  \label{eq:D_dv_3} \\
  &E_{s,t}^{\rm{B}}, n_{s,t}^{\rm{HS}},m_{s,t}^{\rm{ASto}}, \forall s \in \mathbb{S}, \forall t \in \mathbb{T}  \label{eq:D_dv_4} 
\end{align}
\end{subequations}

\subsubsection{The Overall MSEP Model}
Summarizing all the above, the overall optimization model, i.e. MSEP model is established, denoted as
\begin{align}
  \begin{array}{l}
  \mathop {\max }\limits_{\left( \ref{eq:D_dv_1} \right) - \left( \ref{eq:D_dv_4} \right)} \left( {\ref{eq:NPC}} \right)\\
  {\rm{s.t.}} \quad \left( \ref{eq:NPC_inv} \right) - \left( \ref{eq:CER_2} \right)
  \end{array}
  \label{eq:Obj}
\end{align}

The proposed MSEP model (\ref{eq:Obj}) is a typical large-scale linear programming (LP) problem since yearly data with hourly time resolution are utilized at every planning stage, to well describe the intermittence of renewable power generation. Directly solving such a large-scale optimization problem may be inefficient or even infeasible due to numerical troubles. To address the issues above, a Dantzig-Wolfe decomposition (DWD)-based method is introduced in Section \ref{sec:DWD_CG}.

\subsection{Calculation Methods for Performance Indices}
\label{sec:Cal_KPI}

\subsubsection{Levelized Cost of Storage (LCOS)}
LCOS for ESS is defined as the ratio of NPC related to ESS (including investment cost, O\&M cost, charging cost, etc.) to the present value of ESS's discharging energy. However, in such a complex system with strong coupling of electricity, hydrogen, and ammonia, it is difficult to directly calculate the LCOS of BESS, HESS, and AESS. Based on our previous work \cite{yu2023optimal}, internal trading of electricity and hydrogen contributes to evaluate the economic of each part in multi-investor systems.  Therefore, LCOS assessment of BESS, HESS, and AESS is introduced based on virtual internal trading, as shown in Appendix \ref{sec:A1}.

\subsubsection{Levelized Cost of energy (LCOE) for UHVDC}
LCOE for UHVDC is defined as the ratio of $NPC$ to the present value of UHVDC's load, denoted as
\begin{align}
  LCOE = \frac{NPC}{\sum_{s\in \mathbb{S}} \sum_{t \in \mathbb{T}} \delta_{3}(s) P_{s,t}^{\rm{UHVDC}}}
\end{align}
In addition, LCOE for UHVDC can also be determined by the LCOE of wind power, solar power, and thermal power, as well as the LCOS of BESS, HESS, and AESS. Details are presented in Appendix \ref{sec:A1}.

\subsubsection{The Ratio of RES Curtailment $r^{\rm{curt}}$} 
$r^{\rm{curt}}$ is defined as the ratio of curtailment power of RES to the amount of renewable generation over the planning horizon, denoted as
\begin{align}
  r^{\rm{curt}} = \frac{\sum_{s\in \mathbb{S}} \sum_{t \in \mathbb{T}} \delta_{3}(s) P_{s,t}^{\rm{curt}}}{\sum_{s\in \mathbb{S}} \sum_{t \in \mathbb{T}} \delta_{3}(s) \left(P_{s,t}^{\rm{W}}+P_{s,t}^{\rm{S}}\right)} \times 100\%
\end{align}

\subsubsection{The Ratio of CFPP Retirement $r^{\rm{reti}}$}
$r^{\rm{reti}}$ is defined as the ratio of cumulative retired capacities to installed capacities of CFPP, denoted as
\begin{align}
  r^{\rm{reti}} = \frac{\sum_{s\in \mathbb{S}} \Delta C_s^{\rm{CFPP,reti}}}{\sum_{s\in \mathbb{S}} \Delta C_s^{\rm{CFPP}}} \times 100\%
\end{align}

\section{A Combined Dantzig-Wolfe Decomposition (DWD) and Column Generation (CG) Algorithm}
\label{sec:DWD_CG}
In this section, a DWD-CG algorithm is introduced to efficiently solve the proposed MSEP model (\ref{eq:Obj}). 

For the sake of simplicity, slack variables are introduced to convert the inequality constraints into the equivalent equality constraints. Therefore, the proposed MSEP model (\ref{eq:Obj}) can be formulated as follows:
\begin{subequations}
\begin{align}
  &(\mathbf{P}) \quad \min \limits_{\bm{x}} \sum \limits_{i=1}^{S} {\bm{c}}_{i}^{\top}{\bm{x}}_{i} \label{eq:P_1} \\
  &\quad \rm{s.t.} \nonumber \\
  & \quad \quad \; 
  \left[
  \begin{array}{cccc}
    {\bm{B}}_1 &{\bm{B}}_2 &\cdots &{\bm{B}}_S  \\ \hdashline
    {\bm{A}}_1 &{\bm{0}} &\cdots &{\bm{0}}\\
    {\bm{0}} &{\bm{A}}_2 &\cdots &{\bm{0}}\\
    \vdots &\vdots &\ddots &\vdots\\
    {\bm{0}} &{\bm{0}} &\cdots &{\bm{A}}_S
  \end{array}
  \right]
  \begin{bmatrix}
    {\bm{x}}_1  \\
    {\bm{x}}_2  \\
    \vdots \\
    {\bm{x}}_S
  \end{bmatrix}
  =
  \left[
  \begin{array}{c}
    {\bm{h}}_0 \\  \hdashline
    {\bm{h}}_1 \\
    {\bm{h}}_2 \\
    \vdots\\
    {\bm{h}}_S 
  \end{array}
  \right]
   \label{eq:P_2}\\
  & \quad \quad  \quad \quad \;{\bm{x}} \geq \bm{0} \label{eq:P_3}
\end{align}
\end{subequations}

\noindent
where $\bm{x} = \left[{\bm{x}}_1^{\top},{\bm{x}}_2^{\top},\cdots,{\bm{x}}_S^{\top}  \right]^{\top}$. The complex constraints $\sum_{i=1}^{S}\bm{B}_i \bm{x}_i= \bm{h}_0$ are derived from constraints (\ref{eq:Capa_1}) in MSEP model (\ref{eq:Obj}).


Dantzig-Wolfe Decomposition (DWD) is a well-known method to solve large-scale linear programming (LP) with a block-angular structure. DWD is based on the Minkowski theorem \cite{zamfirescu2008minkowski,wirtz20235th}, i.e., \emph{every compact convex set is the convex hull of its set of extreme points}. Therefore, compact convex set $\left\{ \bm{x}_i| \bm{A}_i  \bm{x}_i =  \bm{h}_i\right\}$ can be represented as:
\begin{align}
  \bm{x}_i = \sum\limits_{j=1}^{M_i}\lambda_{ij}\bm{v}_{i}^{(j)} \label{eq:Minkwoski_theorem}
\end{align}

Applying (\ref{eq:Minkwoski_theorem}) to the original problem (\textbf{P}) in (\ref{eq:P_1}), the master problem (MP) is written as follows:
\begin{subequations}
\begin{align}
  &(\mathbf{MP}) \quad f = \min \limits_{\bm{\lambda}} \sum\limits_{i=1}^{S} \sum\limits_{j=1}^{M_i}  \left({\bm{c}}_{i}^{\top} \bm{v}_{i}^{(j)} \lambda_{ij}\right)  \qquad \qquad \quad\label{eq:MP_1}\\
  &\quad \;\rm{s.t.} \nonumber \\
  &\quad \quad \quad \; \sum\limits_{i=1}^{S} \sum\limits_{j=1}^{M_i} \bm{B}_i \bm{v}_{i}^{(j)} \lambda_{ij} = \bm{h}_0 \label{eq:MP_2}\\
  &\quad \quad \quad \;  \sum\limits_{j=1}^{M_i}\lambda_{ij} = 1, \; i =1,2,\ldots,S \label{eq:MP_3}\\
  &\quad \quad \quad \; {\bm{\lambda}} \geq \bm{0} \label{eq:MP_4}
\end{align}
\end{subequations}
\noindent
where $\bm{\lambda} = \left[{\bm{\lambda}_1}^{\top},\ldots,{\bm{\lambda}_i}^{\top},\ldots,{\bm{\lambda}_S}^{\top}\right]^{\top}$, and ${\bm{\lambda}_i}$ represents the convex combination coefficients of the $i$th subproblem, denoted as ${\bm{\lambda}_i} = \left[ {\lambda_{i1}},\ldots,{\lambda_{iM_i}}\right]^{\top}$, with the convex constraints in (\ref{eq:MP_3}) and non-negativity constraints in (\ref{eq:MP_4}). Furthermore, problem ({\bf{MP}}) can be written in matrix form as follows:
\begin{subequations}
  \begin{align}
    &(\mathbf{MMP}) \quad f = \min \limits_{\bm{\lambda}} ({\bm{CV}})^{\top} \bm{\lambda}  \qquad \qquad \qquad \quad\label{eq:MMP_1}\\
    &\quad \;\rm{s.t.} \nonumber \\
    &\quad \quad \quad \; ({\bm{BV}})\bm{\lambda} = \bm{h}_0 \label{eq:MMP_2}\\
    &\quad \quad \quad \;  {\bm{D}}\bm{\lambda} = \bm{1}_{S \times 1} \label{eq:MMP_3}\\
    &\quad \quad \quad \; {\bm{\lambda}} \geq \bm{0} \label{eq:MMP_4}
  \end{align}
\end{subequations}
\noindent
where $\bm{CV} = \left[ \bm{c}_1^{\top}\bm{V}_1,\ldots,\bm{c}_i^{\top}\bm{V}_i,\ldots,\bm{c}_S^{\top}\bm{V}_S\right]^{\top}$. $\bm{V}_i$ is the set of extreme points of the $i$th subproblem, denoted as $\bm{V}_i = \left[ \bm{v}_i^{(1)},\ldots,\bm{v}_i^{(M_i)}\right]$. And $\bm{BV} = \left[ \bm{B}_1 \bm{V}_1,\ldots,\bm{B}_S \bm{V}_S\right]$, $\bm{D} = {\rm{diag}}\left[\bm{1}_{M_1 \times 1},\ldots,\bm{1}_{M_S \times 1}\right]$. Note that $\rm{diag}\left[\cdot \right]$ means the block diagonal matrix.

The dual problem of MMP is denoted as follows:
\begin{subequations}
\begin{align}
  &(\mathbf{D-MMP}) \quad \max \limits_{\bm{\alpha},\bm{\beta}}  {\bm{1}}^{\top}_{S \times 1} \bm{\alpha} + {\bm{h}_0}^{\top} \bm{\beta}  \qquad \qquad \quad\label{eq:DMP_1}\\
  &\quad \;\rm{s.t.} \nonumber \\
  &\quad \quad \quad \;{\bm{D}}^{\top} \bm{\alpha} + (\bm{BV})^{\top} \bm{\beta}   \leq \bm{CV} \label{eq:DMP_2}
\end{align}
\end{subequations}
\noindent
where $\bm{\alpha} = \left[\alpha_{1},\ldots,\alpha_{S}\right]^{\top}$, $\bm{\beta} = \left[\beta_{1},\ldots, \beta_{\left| \bm{h}_0\right|}\right]^{\top}$ are the solution of problem ({\bf{D-MMP}}), which represent the shadow prices (also known as Lagrange multipliers) of constraints (\ref{eq:MMP_3}) and  (\ref{eq:MMP_2}) in problem ({\bf{MP}}), respectively.

According to \cite{mcnamara2015hierarchical}, the reduced cost of the $i$th subproblem (SP) is given in (\ref{eq:SP_1}), which is used as the objective function. Therefore, the problem $(\mathbf{SP}_i)$  is written as follows:
\begin{subequations}
\begin{align}
  & (\mathbf{SP}_i) \quad \phi_i = \min \limits_{\bm{x}_i} \left(\bm{c}_i - \bm{B}_i^{\top} \bm{\beta}\right)^{\top} {\bm{x}_i} - \alpha_{i} \qquad \qquad \quad\label{eq:SP_1}\\
  &\quad \; \; \rm{s.t.} \nonumber \\
  &\quad \quad \quad \;  \bm{A}_i \bm{x}_{i} = \bm{h}_i \label{eq:SP_2}\\
  &\quad \quad \quad \;  \bm{x}_{i} \geq \bm{0} \label{eq:SP_3}
\end{align}
\end{subequations}
\noindent

Since the problem $(\mathbf{SP}_i)$ is solved, if the optimal objective value $ \phi_i <0$, it means that $\bm{x}_i$ is capable of reducing the cost of MP, and so the corresponding $\bm{x}_i$ will be added to $\bm{V}_i$ as a new extreme point. The above step is known as the process of column generation (CG). In addition, the problems $(\mathbf{SP})$ can be solved in parallel to reduce the computational time cost. 

However, it is inefficient to terminate the iteration of DWD-CG until all  $ \phi_i \geq 0$ for large-scale LP. \cite{bazaraa2011linear} provides the termination criteria in (\ref{eq:TC_1}) using upper and lower bounds (\ref{eq:TC_2})--(\ref{eq:TC_3}), denoted as follows:
\begin{subequations}
\begin{align}
  &gap = \left|  \frac{UB-LB}{UB} \right|  < \epsilon    \label{eq:TC_1} \\
  &UB = f  \label{eq:TC_2} \\
  &LB = \bm{\beta}^{\top}\bm{h}_0 + \sum\limits_{i=1}^{S}(\phi _i + \alpha_i) \label{eq:TC_3} 
\end{align}
\end{subequations}
\noindent
where $\epsilon $ is the given error tolerance. 

\begin{figure}[t]
  \centering
  \includegraphics[width=3.46in]{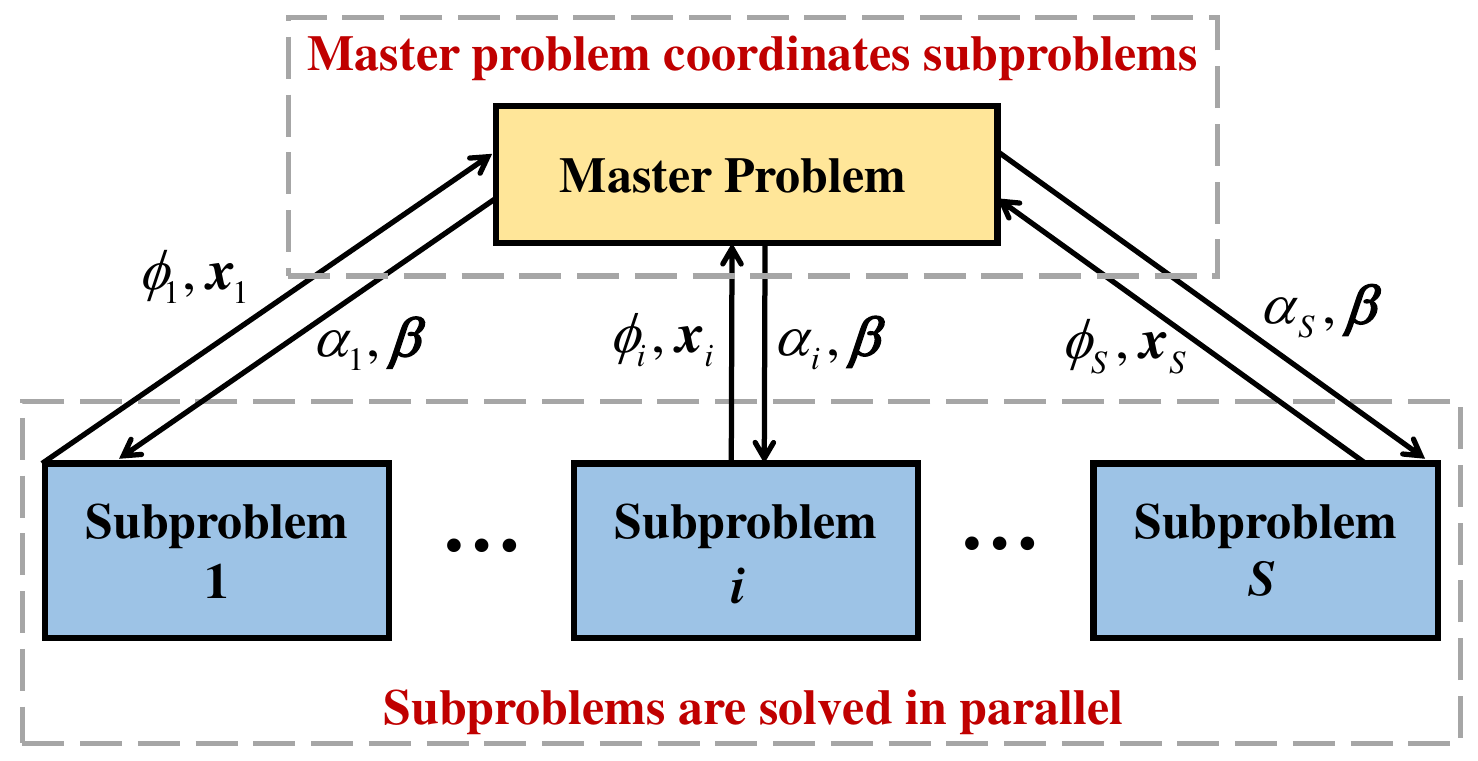}
  \caption{Relationship between master problem and several subproblems}
  \label{fig:DWD_CG_MP_SP}
\end{figure}

Therefore, relationship between MP and $S$ SP are summarized in Fig. \ref{fig:DWD_CG_MP_SP}. Furthermore, the entire procedure of the proposed DWD-CG algorithm is presented in Algorithm \ref{alg:DWD_CG}.

\begin{algorithm}[t]
  \caption{ A Combined Dantzig-Wolfe Decomposition (DWD) and Column Generation (CG) Algorithm}
  \label{alg:DWD_CG}
  \LinesNumbered
  \KwIn{Set $\bm{V}_i := \left[ \bm{v}_i^{(1)}\right],\forall i \in \mathbb{S}$, $\epsilon  := 10^{-4}$, $K := 100$}
  \KwOut{Optimal solution $obj$ and $\bm{x}_i^*,\forall i \in \mathbb{S}$}
  \For{$k\leftarrow 1$ \KwTo $K$}{
      Solve the problem (\textbf{MP}) to obtain the convex combination coefficients $\bm{\lambda}^{(k)}$\;
      Solve the problem (\textbf{D-MMP}) to obtain the shadow prices  $\bm{\alpha}^{(k)}$ and $\bm{\beta}^{(k)}$\;
      Solve the subproblem $(\mathbf{SP}_i),\forall i \in \mathbb{S}$ in parallel to obtain the optimal solution: $\left(\bm{x}_i^{(k)}, \phi _i^{(k)}\right),\forall i \in \mathbb{S}$\;
      \For{$i\leftarrow 1$ \KwTo $S$}{
        \If{ $ \phi_i < 0$}{Add a new extreme point to the set, i.e., $\bm{V}_i = \left[\bm{V}_i, \bm{x}_i^{(k)}\right]$\;}
      }
      Calculate the current gap $gap^{(k)}$ according to (\ref{eq:TC_1})--(\ref{eq:TC_3})\;
      \If{$gap^{(k)}<\epsilon$ }{
        Calculate the optimal solution: $\bm{x}_i^* = \sum\limits_{j=1}^{M_i}\lambda_{ij}^{(k)}\bm{v}_{i}^{(j)},\forall i \in \mathbb{S}$\; 
        Calculate the optimal objective: $obj = f^{(k)}$ \;
        Terminate the algorithm\;}
  }
\end{algorithm}

\emph{Remark1:} In Algorithm \ref{alg:DWD_CG}, note that the initialization of sets of extreme points $\bm{V}_i$ should be satisfied to complex constraints in (\ref{eq:MP_2}), and corresponding objective value $f$ in (\ref{eq:MP_1}) should be large enough to guarantee a reliable upper bound.

\section{Case Studies}
\label{sec:case}
In this section, case studies are performed using data from a real-life system, i.e.,the  UHVDC project from Gansu province to Shandong province in China. First, the optimal planning results of the MSEP model are presented with a detailed discussion. Second, the techno-economic analysis of BESS, HESS, and AESS is discussed. Then, the performances of the proposed DWD-CG algorithm for solving the MSEP model are presented compared to directly solving by \emph{Gurobi}. Finally, sensitivity analysis of carbon emission reduction targets is discussed.

\subsection{Case Description and Setup}
\label{sec:descrip}
To study the proposed method, the MSEP model in Section \ref{sec:MSEP}, and corresponding solution approaches in Section \ref{sec:DWD_CG} are established in \emph{MATLAB R2023a} and solved by \emph{Gurobi 10.0.0}, environment on a desktop computer with Intel(R) Core (TM) i7-10700 CPU @ 2.90GHz processor with 128 GB RAM.

The real-life planning UHVDC project from Gansu to Shandong in China \cite{GSUHVDC} is used in the case studies, and the load curves of UHVDC during different seasons are shown in Fig. \ref{fig:System_Topology}. Wind and solar power are generated based on the real historical meteorology data from the project, full load hours (FLH) of wind power is 3000 hours, and that of solar power is 1500 hours. Yearly data with hourly time resolution are utilized for each planning stage, i.e., $\Delta T = 1$h, $ N = 8760$h. The carbon emission reduction target $r_s^{\rm{CER}}$ for each stage is assumed to decrease linearly from $0\%$ to $r_S^{\rm{CER}}$ during planning stages. The investment and operation parameters of WT, PV, BESS, AE, HS, PEMFC, and ASyn, can be found in our previous work \cite{yu2023optimal,yu2023optimal_Iso}. Moreover, parameters related to CFPP and ASto are listed as follows.

CFPP \cite{RuleESS2023}: $I_{1}^{\rm{CFPP,init}} = 3,500 \ \rm{RMB/kW}$, $I_{1}^{\rm{CFPP,reti}} = 500 \ \rm{RMB/kW}$, $I^{\rm{CFPP,O\&M}} = 3\%$, $Y_{\rm{CFPP}} = 25$ years, $\gamma _{\rm{P2C}} = 0.300 \ \rm{kg/kWh}$, $\underline{\eta}^{\rm{CFPP}} = 30\%$, $\overline{\eta}^{\rm{CFPP}} = 100\%$, and $\mu _{\rm{CF}} = 0.738 \ \rm{kg/kWh}$.

ASto: $I_{1}^{\rm{ASto,init}} = 5,500 \ \rm{RMB/t}$, $I^{\rm{ASto,O\&M}} = 1\%$, $Y^{\rm{ASto}} = 20$, $\underline{\eta}^{\rm{ASto}} = 10\%$, and $\overline{\eta}^{\rm{ASto}} = 90\%$.

Green hydrogen price $\pi_s^{\rm{H,purch}}$ for each stage is assumed decreasing linearly from $2.0 \ \rm{RMB/Nm^3}$ to $1.5 \ \rm{RMB/Nm^3}$ during planning stages, and so is the green ammonia price $\pi_S^{\rm{A,purch}}$, decreasing linearly from $4,000 \ \rm{RMB/Nm^3}$ to $3,500 \ \rm{RMB/t}$. Since coal-fired generation is considered, the produced hydrogen and ammonia can only compete with gray hydrogen and ammonia. Therefore, the selling prices are set as $\pi^{\rm{H,sell}} = 1.2 \ \rm{RMB/Nm^3}$ and $\pi^{\rm{A,sell}} = 3,000 \ \rm{RMB/t}$, which are less than that of green hydrogen and ammonia.

In addition, the maximum capacities of WT, PV, CFPP, BESS, AE, HS, FC, ASyn, ASto are set as $24.20\rm{GW}$, $32.03\rm{GW}$, $8.00\rm{GW}$, $45,000\rm{GWh}$, $8.00\rm{GW}$, $50.00\rm{MNm^3}$, $8.00\rm{GW}$, $10.00\rm{Mt}$, and $10.00\rm{Mt}$, respectively.



\subsection{Case Study of a 5-Stage Planning}
In this section, 5-stage planning toward carbon neutrality is studied, i.e., $S=5$, $\Delta S=3$ years, and $r_S^{\rm{CER}} = 100\%$. To well demonstrate different values of BESS, HESS, and AESS, four cases are introduced for comparison:

\textbf{Case1}: Only BESS is used to realize the carbon-free target, i.e., $C_s^{j} = 0, \forall s \in \mathbb{S}, \forall j \in \Omega_{\rm{H}} \bigcup \Omega_{\rm{A}}$.

\textbf{Case2}: Both BESS and HESS are used to realize the carbon-free target, i.e., $C_s^{j} = 0, \forall s \in \mathbb{S}, \forall j \in \Omega_{\rm{A}}$.

\textbf{Case3}: BESS, HESS  and AESS are all used to realize the carbon-free target, but hydrogen/ammonia trading with external markets are not allowed, i.e., $q_s^{\rm{H,purch}} =  0, q_s^{\rm{H,sell}} =  0, q_s^{\rm{A,purch}} =  0,q_s^{\rm{A,sell}} =  0, \forall s \in \mathbb{S}, \forall t \in \mathbb{T}$.

\textbf{Case4}: BESS, HESS, and AESS are all used to realize carbon free target, so is the external trading of hydrogen/ammonia.

Corresponding optimal planning results and performance indices for Case1--Case4 are listed in Table \ref{tab:Performance_Cases}.

Specifically, in Case1, the ratio of RES curtailment exceeds $43.62\%$, which reflects the mode of using base-load of RES to fit the load demand of UHVDC when only BESS can be used. Furthermore, a large capacity of BESS is needed; approximately 38h of electricity storage demand is derived even using the capacity of UHVDC (8GW) as a benchmark. As a result, high levels of LCOE and LCOS are reached with the value of $0.5913 \ \rm{RMB/kWh}$ and $3.1364 \ \rm{RMB/kWh}$, respectively.

In Case2, since the HESS is introduced, the curtailment rate and capacity of BESS are decreased, which leads to a reduction of $12.28\%$ in LCOE compared to Case1.

In Case3, there is an order of magnitude reduction in the optimal capacity of BESS, i.e., only $22.62 \ \rm{GWh}$ BESS is needed. The curtailment rate of RES also decreases to $12.77\%$, a reduction of $70.72\%$ compared to Case1. Therefore, LCOE in Case3 reduces to $0.4324 \ \rm{RMB/kWh}$. 

In Case4, a slight reduction in LCOE is reached compared to Case3, which is approximately $1.272\%$. This is because more capacity of AE is needed, leading to the largest installed capacities of WT and PV. It is easier to meet the load demand of UHVDC. 

Following the results in Case3 and Case4, we find that $LCOS_{\rm{B}} < LCOS_{\rm{H}} < LCOS_{\rm{A}}$. It seems that HESS and AESS are not necessary since they are much more costly than BESS. However, comparing Case1--Case4 together, we can find that $LCOE_{\rm{G}}$ and $LCOS_{\rm{B}}$ are lowered since HESS and AESS are introduced. Part of CFPP is preserved when AESS can generate using ammonia-fired in CFPP units. Especially in Case3, just $34.27\%$ CFPP units are retired, which significantly increases the utilization rate of CFPP facilities and avoids a waste of CFPP investment due to carbon-free targets.

In other words, the value of HESS and AESS is not reflected in a low LCOS of themselves but reflected in the contribution to lowering the curtailment rate of RES and capacity of BESS, so that $LCOE_{\rm{G}}$ and $LCOS_{\rm{B}}$ are reduced, leading to a lower LCOE for the whole system.

\begin{table*}[t]\scriptsize
  \renewcommand{\arraystretch}{2.0}
  \caption{Performance Comparison of Different Cases}
  \label{tab:Performance_Cases}
  \centering
  \resizebox{1.0\linewidth}{!}{
  \begin{threeparttable}[b]
  \begin{tabular}{cccccccccccc}
  \hline \hline 
  \multirow{2}{*}{Case}  &\multirow{2}{*}{\tabincell{c}{Optimal sizes \tnote{a} \\(GW,GW,GW,GW,GWh, \\GW,$\rm{MNm^3}$,GW,Mt,Mt)}}   &\multirow{2}{*}{\tabincell{c}{$r^{\rm{curt}}$ \\(\%)}} &\multirow{2}{*}{\tabincell{c}{$r^{\rm{reti}}$  \\(\%)}}  &\multirow{2}{*}{\tabincell{c}{${LCOE}$ \\(RMB/kWh)}}   &\multirow{2}{*}{\tabincell{c}{${LCOE}_{\rm{G}}$ \\(RMB/kWh)}} &\multicolumn{2}{c}{BESS} &\multicolumn{2}{c}{HESS} &\multicolumn{2}{c}{AESS} \\
  \cmidrule(r){7-8}  \cmidrule(r){9-10} \cmidrule(r){11-12} 
  &{} &{}  &{}  &{}  &{} &\tabincell{c}{${LCOS}_{\rm{B}}$ \\(RMB/kWh)} &\tabincell{c}{$r_{\rm{B}}^{\rm{CAPEX}}$ \tnote{b}\\(\%)}  
  &\tabincell{c}{${LCOS}_{\rm{H}}$ \\(RMB/kWh)} &\tabincell{c}{$r_{\rm{H}}^{\rm{CAPEX}}$ \tnote{d} \\(\%)} 
   &\tabincell{c}{${LCOS}_{\rm{A}}$ \\(RMB/kWh)} &\tabincell{c}{$r_{\rm{A}}^{\rm{CAPEX}}  $\tnote{d} \\(\%)}  \\
  \hline
  Case1 &\tabincell{c}{$\left\{ 20.70,32.03,4.77,4.77,298.66, \right.$ \\$\left. 0.00,0.00,0.00,0.00,0.00 \right\}$}  &$43.62$ &$100.00$ &$0.5913$  &$0.3523$  &$3.1364$  &$87.56$  &$\rm{N/A}$ &$\rm{N/A}$  &$\rm{N/A}$ &$\rm{N/A}$  \\
  Case2 &\tabincell{c}{$\left\{ 20.70,32.03,4.39,4.39,219.71, \right.$ \\$\left. 1.73,50.00,1.32,0.00,0.00 \right\}$} &$32.96$ &$100.00$ &$0.5187$  &$0.3081$  &$2.7753$  &$87.70$   &$2.2403$ &$54.15$  &$\rm{N/A}$ &$\rm{N/A}$  \\
  Case3 &\tabincell{c}{$\left\{ 18.55,30.50,4.39,1.50,22.62, \right.$ \\$\left. 8.67,50.00,1.14,4.86,0.73 \right\}$}  &$12.77$ &$34.27$ &$0.4324$  &$0.2614$  &$1.2526$  &$76.88$   &$1.4767$ &$41.00$  &$2.2456$ &$9.46$  \\
  Case4 &\tabincell{c}{$\left\{ 24.2,32.03,4.38,2.71,31.06, \right.$ \\$\left. 10.67,50.00,1.88,10,0.85\right\}$} &$11.89$ &$61.83$  &$0.4269$  &$0.2559$  &$1.2245$  &$76.84$   &$1.3139$ &$31.61$  &$3.0783$ &$16.61$  \\

  \hline \hline
  \end{tabular}
  \begin{tablenotes}
  \footnotesize
  \item[a] Cumulative capacities of WT, PV, installed CFPP, retired CFPP, BESS, AE, HS, FC, ASyn, and ASto, respectively.
  \item[b] The ratio of the cost related to capital expenditure (CAPEX) including initial investment and fixed O\&M.
  \end{tablenotes}
  \end{threeparttable}
  }
\end{table*}



\subsection{Typical Operation Mode Analysis for BESS, HESS, and AESS}

\begin{figure}[t]
  \centering
  \includegraphics[width=3.46in]{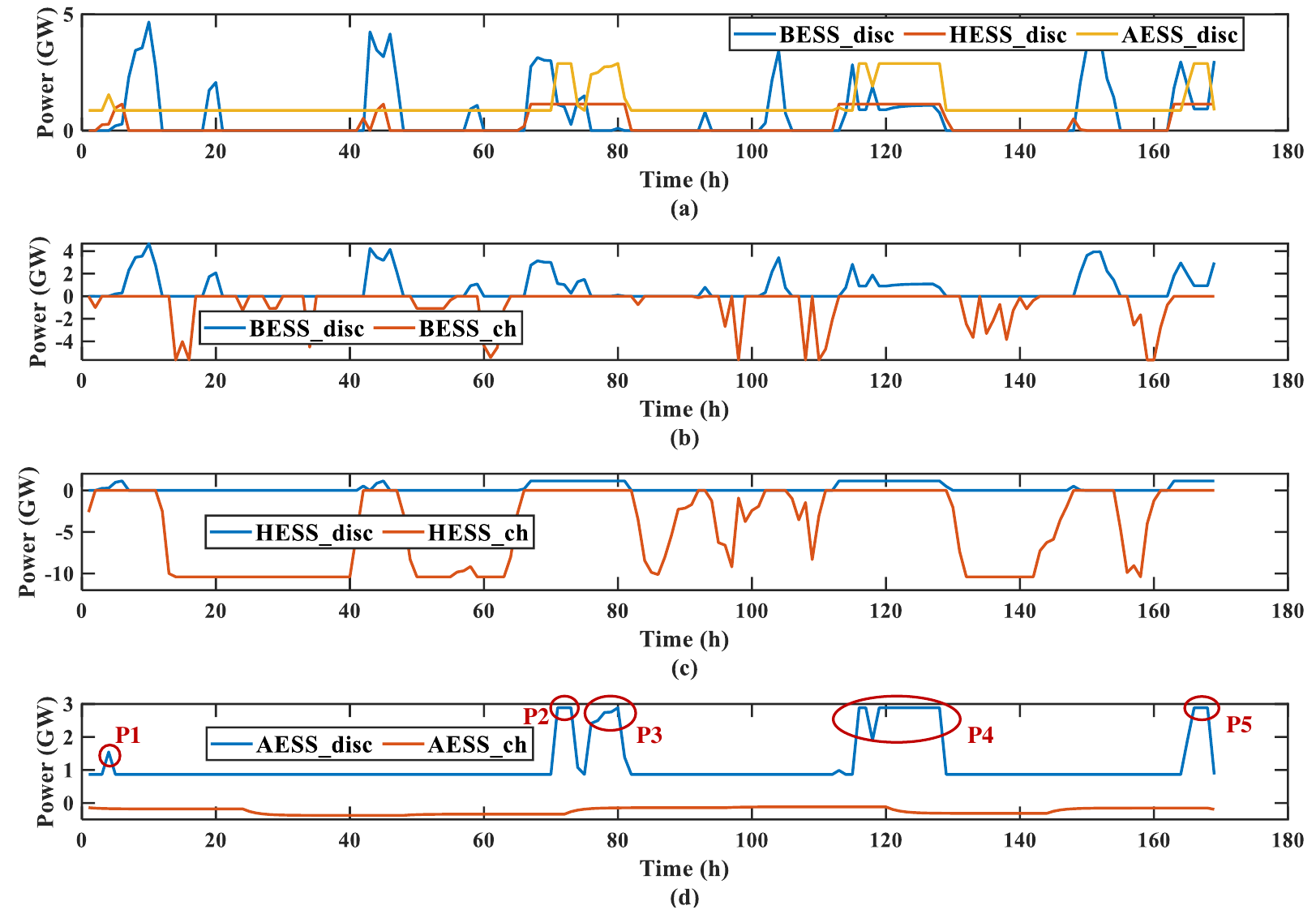}
  \caption{Weekly operation of multiple energy storage systems at the final stage. (a) Discharging profiles of BESS, HESS, and AESS. (b) Discharging and charging profiles of BESS. (c) Discharging and charging profiles of HESS. (d) Discharging and charging profiles of AESS.}
  \label{fig:ESS_Profiles}
\end{figure}

\begin{figure}[t]
  \centering
  \includegraphics[width=3.46in]{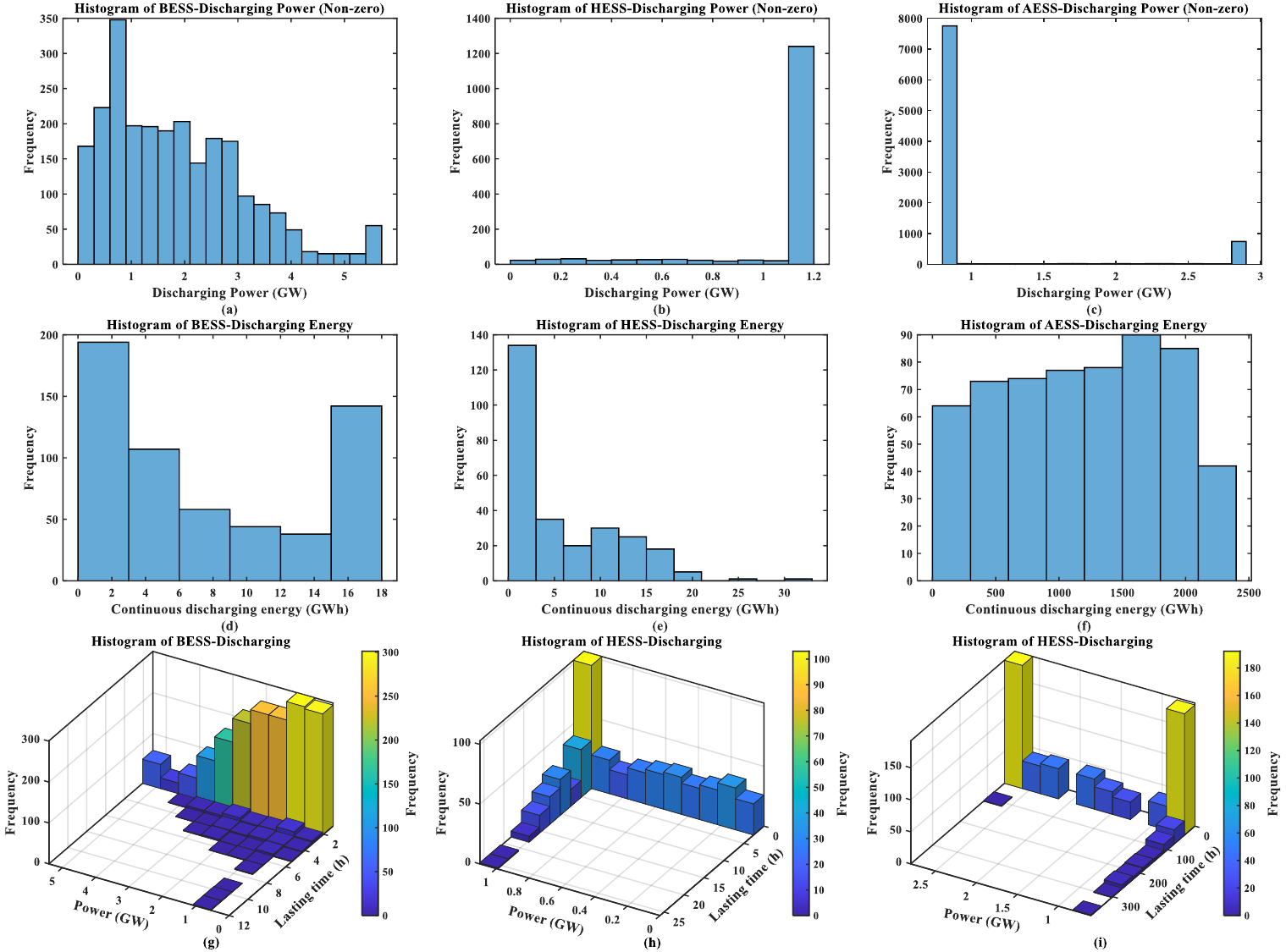}
  \caption{Discharging behaviors of multiple energy storages. (a) Distribution of BESS discharging power. (b) Distribution of HESS discharging power. (c) Distribution of AESS discharging power. (d) Distribution of BESS continuous discharging energy. (e) Distribution of HESS continuous discharging energy. (f) Distribution of AESS continuous discharging energy. (g) Joint distribution of BESS discharging power and lasting time. (h) Joint distribution of HESS discharging power and lasting time. (i) Joint distribution of AESS discharging power and lasting time.}
  \label{fig:Histogram_MESS}
\end{figure}

To further demonstrate the different roles of BESS, HESS, and AESS, the yearly operation of the proposed system at the final stage in Case3 is studied. First, charging/discharging power curves of BESS, HESS, and AESS in February are plotted in Fig. \ref{fig:ESS_Profiles}. The different discharging behaviors of BESS, HESS, and AESS are counted and plotted in Fig. \ref{fig:Histogram_MESS}. 

Specially, from Fig. \ref{fig:ESS_Profiles} (b), we find that the maximum charging and discharging power of BESS are the same. The power of BESS is widely distributed in multiple scenarios since BESS mainly handles the volatility of RES, which is clearly shown in Fig. \ref{fig:Histogram_MESS} (a).

From Fig. \ref{fig:ESS_Profiles} (c), HESS shows a higher capacity for charging than that for discharging since HESS also provides hydrogen for AESS. Discharging of HESS almost concentrates on the full load of PEMFC, shown in \ref{fig:Histogram_MESS} (b).

On the contrary, AESS exhibits a higher capacity for discharging than that for charging, as shown in \ref{fig:ESS_Profiles} (d). Discharging power of AESS shows a bipolar distribution in \ref{fig:Histogram_MESS} (c), which matches the operational characteristics of CFPP, i.e., the base-load operation is maintained most of the time, and full load operation occurs very rarely. Specifically, there are five periods $P_1$--$P_5$ when ammonia-fired generation under full load during the selected week shown in \ref{fig:ESS_Profiles} (d).

Furthermore, comparing Fig. \ref{fig:Histogram_MESS} (d)--(f) together, the results indicate that the single continuous discharging energy for BESS and HESS is the same order of magnitude, but that of AESS is nearly two orders of magnitude higher. A similar situation can be found in \ref{fig:Histogram_MESS} (g)--(i). Therefore, it is suitable for HESS and AESS to handle the intermittence of RES with long periods but low probability.

\begin{figure}[t]
  \centering
  \includegraphics[width=3.46in]{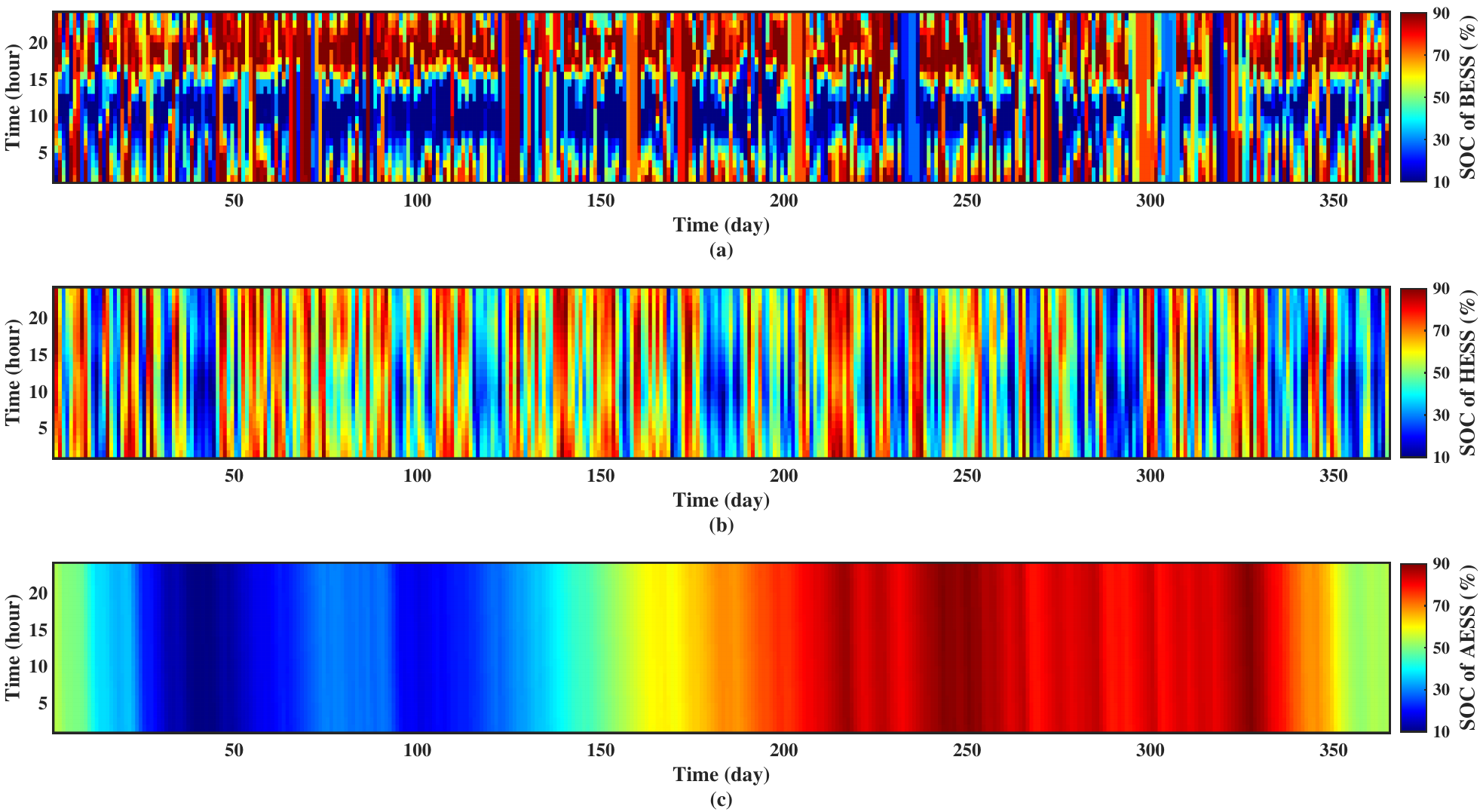}
  \caption{Yearly operation of multiple energy storage systems. (a) SOC of BESS. (b) SOC of HESS. (c) SOC of AESS.}
  \label{fig:ESS_SOC}
\end{figure}

In addition, SOC of BESS, HESS, and AESS is presented in Fig. \ref{fig:ESS_SOC}. The results show that the SOC of BESS varies within the day, while cross-day variation is uncommon. The SOC of HESS changes over multiple days, and that of AESS features monthly and seasonal variations. Therefore, HESS and AESS are also known as seasonal energy storage \cite{brey2021use}.

\subsection{The Advantage of DWD-CG Solution Approach}
\label{sec:case_DWD_CG}
To demonstrate the advantage of the proposed DWD-CG method, Case3\_S\_5--Case3\_S\_30 is studied, whose setting is the same as Case3, but the scale of the problem, i.e., $S$ and $\Delta S$ are changed, listed in Table \ref{tab:Performance_DWD_CG} with the number of variables and constraints. Two methods are utilized for comparison: directly solving the MSEP model (\ref{eq:Obj}) by \emph{gurobi} \cite{gurobi} and the proposed DWD-CG method. Two indices are introduced, i.e., objective value and computational CPU times, representing accuracy and efficiency, respectively. Furthermore, the gaps of optimal objective values between two methods are listed in Table \ref{tab:Performance_DWD_CG}.

The results in Table \ref{tab:Performance_DWD_CG} indicate that two methods can obtain an optimal solution when the scale of the problem is relatively small, as in Case3\_S\_5--Case3\_S\_10. The time costs of two methods are on the same order of magnitude, nearly thousands of seconds. Furthermore, the gap between two methods is approximately $10^{-4}$ and below, demonstrating the accuracy of the proposed DWD-CG method. 

However, when the scale is further increased, such as in Case3\_S\_15--Case3\_S\_30, \emph{Gurobi} cannot obtain an optimal or feasible solution within the given time limit. In some cases, numerical trouble is encountered, leading to the infeasibility of obtaining a solution as in Case3\_S\_25.

On the contrary, the proposed DWD-CG method can still calculate the optimal solutions for Case3\_S\_15--Case3\_S\_30. The time cost in Case3\_S\_30 is only $19,646$ s, which is far less than the given time limit of $36,000$ s. This reveals that the proposed DWD-CG method is efficient for solving large-scale optimization problems.

\begin{figure}[t]
  \centering
  \includegraphics[width=3.46in]{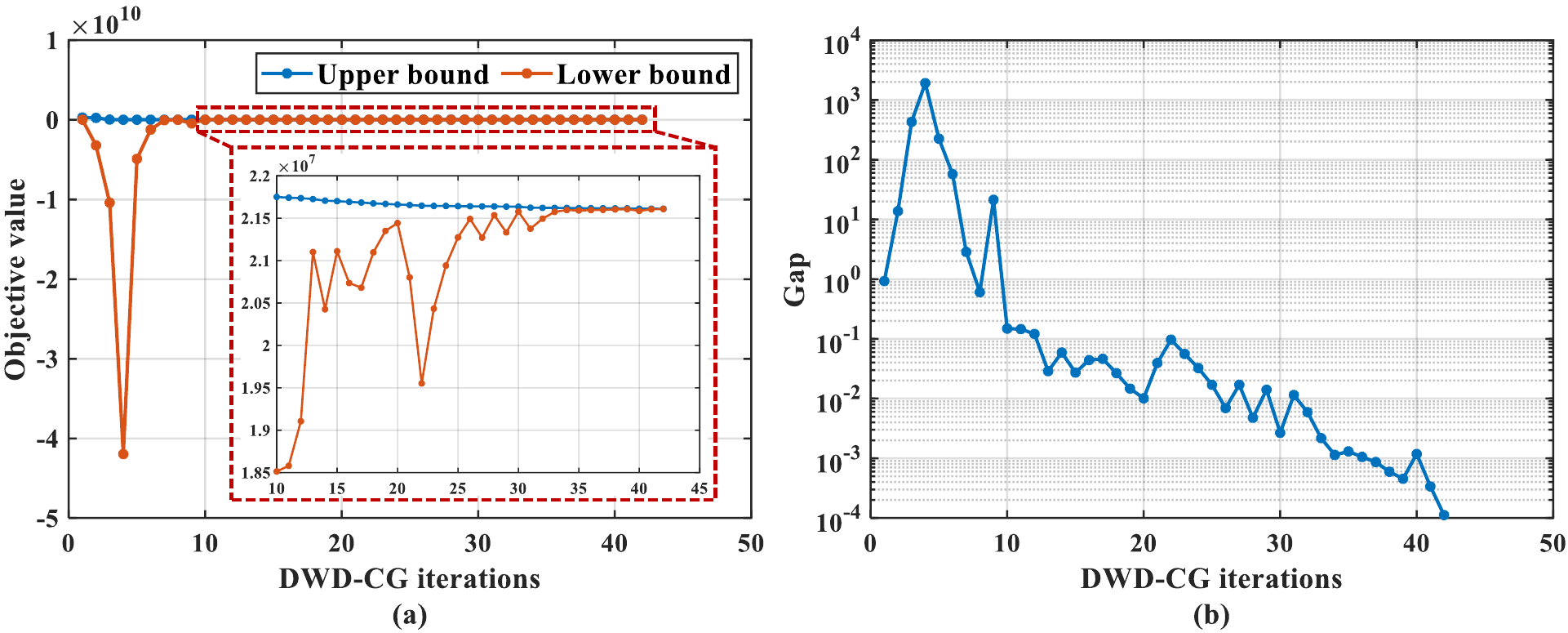}
  \caption{Convergence of the proposed DWD-CG in Case3\_S\_30. (a) Evolution of the upper and lower bounds. (b) Evolution of the gap in iteration process.}
  \label{fig:DWD_CG_Iterations}
\end{figure}

Furthermore, the upper and lower bound defined in (\ref{eq:TC_2})--(\ref{eq:TC_3}) during the iteration process in Case3\_S\_30 are plotted in Fig. \ref{fig:DWD_CG_Iterations} (a), and the corresponding gaps defined in (\ref{eq:TC_1}) are shown in Fig. \ref{fig:DWD_CG_Iterations} (b). Based on the results in Fig. \ref{fig:DWD_CG_Iterations} (b), we can sacrifice the accuracy of the solution to reduce the computational burden, i.e., increasing the error tolerance $\epsilon $. Therefore, DWD-CG has a controllable computational burden in solving large-scale optimization problems, which indicates that DWD has the potential to solve larger-scale problems.

\begin{table*}[t]\scriptsize
  \renewcommand{\arraystretch}{2.0}
  \caption{Performance Comparison of Different Methods for Solving large-scales MSEP Model}
  \label{tab:Performance_DWD_CG}
  \centering
  \begin{threeparttable}[b]
  \begin{tabular}{ccccccccccc}
  \hline \hline 
  \multirow{2}{*}{Case}  &\multirow{2}{*}{$S$} &\multirow{2}{*}{\tabincell{c}{$\Delta S$ \\(year)}}  & \multirow{2}{*}{\tabincell{c}{Number of \\Variables}} &\multirow{2}{*}{\tabincell{c}{Number of \\Constraints}} &\multicolumn{2}{c}{Gurobi} &\multicolumn{3}{c}{DWD-CG} &\multirow{2}{*}{Gap}\\
  \cmidrule(r){6-7}  \cmidrule(r){8-10} 
  &{} &{} &{}  &{} &\tabincell{c}{Objective value \\(billion RMB)} &\tabincell{c}{CPU \\times (s)} &\tabincell{c}{Objective value \\(billion RMB)} &\tabincell{c}{CPU \\times (s)} &\tabincell{c}{Iterations}\\
  \hline
  Case3\_S\_5 &$5$ &$1$ &$1,009,295$ &$2,547,800$ &$82.2151$ &$1,270$ &$82.1828$ &$3,169$  &$20  $ &$3.929 \times 10^{-4}$ \\
  Case3\_S\_10 &$10$ &$1$  &$2,018,590$ &$5,095,600$ &$151.5809$ &$3,740$ &$151.5820$ &$4,650$ &$22$ &$7.257 \times 10^{-6}$ \\
  Case3\_S\_15 &$15$ &$1$  &$3,027,885$ &$7,643,400$ &$\rm{N/A}$\tnote{a} &$36,000$ &$189.9779$ &$8,843$ &$30$ &$\rm{N/A}$\tnote{b} \\
  Case3\_S\_20 &$20$ &$1$  &$4,037,180$ &$10,191,200$ &$\rm{N/A}$\tnote{a} &$36,000$ &$203.8751$ &$11,695$ &$36$ &$\rm{N/A}$\tnote{b} \\
  Case3\_S\_25 &$25$ &$1$  &$5,046,475$ &$12,739,000$ &$\rm{N/A}$\tnote{c} &$36,000$ &$215.3415$ &$16,175$ &$41$ &$\rm{N/A}$\tnote{b} \\
  Case3\_S\_30 &$30$ &$1$  &$6,055,770$ &$15,286,800$ &$\rm{N/A}$\tnote{a} &$36,000$ &$216.1162$ &$19,646$ &$42$ &$\rm{N/A}$\tnote{b} \\

  \hline \hline
  \end{tabular}
  \begin{tablenotes}
  \footnotesize
  \item[a] No optimal or suboptimal solution is obtained within the time limit (set as 36,000 seconds, i.e., 10 hours).
  \item[b] Gap between \emph{Gurobi} and DWD-CG can not be calculated since only the proposed DWD-CG can obtained an optimal solution.
  \item[c] Numerical trouble encountered.
  \end{tablenotes}
  \end{threeparttable}
\end{table*}

\subsection{Sensitivity Analysis of Carbon Emission Reduction Target}
\label{sec:case_sensitivity_analysis}

\begin{figure}[t]
  \centering
  \includegraphics[width=3.46in]{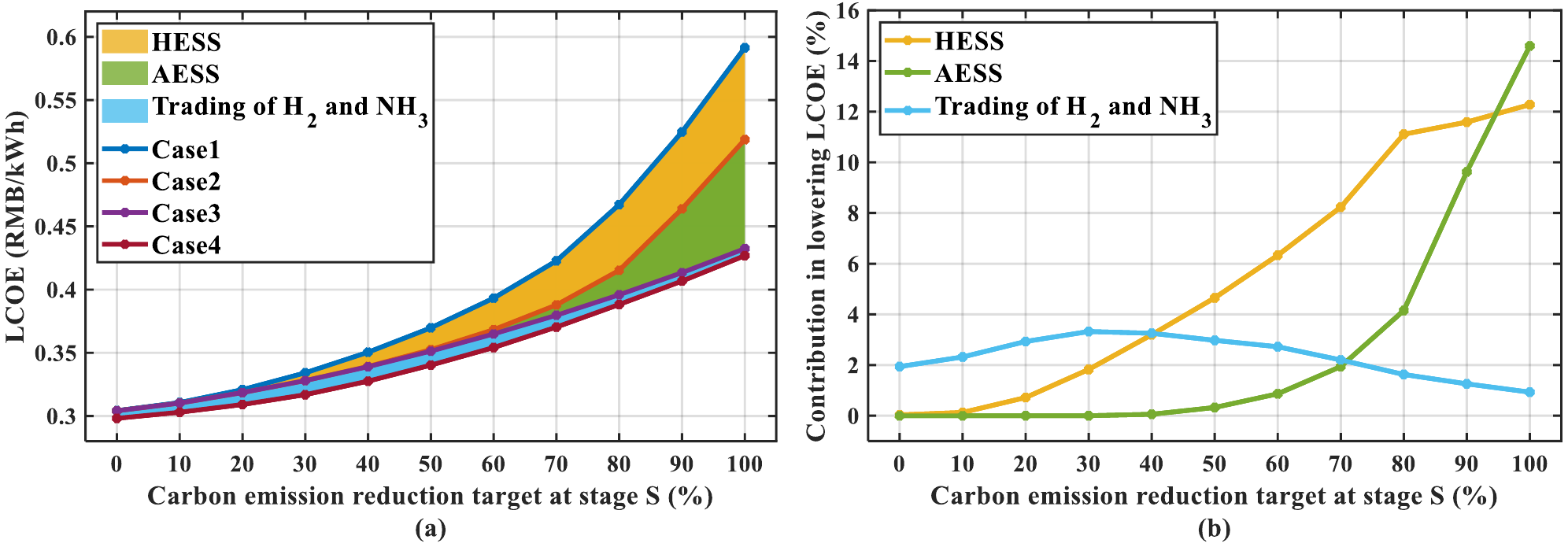}
  \caption{Sensitivity analysis of carbon emission reduction targets. (a) Evolution of LCOE under different values of carbon emission reduction targets. (b) Evolution of the contribution in lowering LCOE for HESS, AESS, and external trading .}
  \label{fig:Sensitivity_CER}
\end{figure}

$r_S^{\rm{CER}}$ is set from $0\%$ to $100\%$ with steps of $10\%$, Case1--Case4 are calculated repeatedly under different value of $r_S^{\rm{CER}}$, and corresponding $LCOE$ are recorded and plotted in Fig. \ref{fig:Sensitivity_CER} (a). The results show that more ambitious carbon reduction targets lead to larger LCOE. Furthermore, the orange region between Case1 and Case2, the green region between Case2 and Case3, and the blue region between Case3 and Case4, represent the roles in lowering LCOE for HESS, AESS, and external trading, respectively. Using the LCOE of Case1 as the denominator, the contribution of HESS, AESS, and external trading in reducing the LCOE is quantified, as shown in Fig. \ref{fig:Sensitivity_CER} (b).

The results show that external trading of hydrogen and ammonia can roughly reduce $2\%$ of LCOE under different values of $r_S^{\rm{CER}}$. However, the influence of $r_S^{\rm{CER}}$ for HESS and AESS is more complicated. Specially discussed as follows:

\emph{a)} When $0\%<r_S^{\rm{CER}} < 10\%$, both HESS and AESS have no contribution in lowering LCOE.

\emph{b)} When $10\%<r_S^{\rm{CER}} < 40\%$, HESS exhibits a positive impact on reducing LCOE, while AESS is still makes a small contribution in lowering LCOE.

\emph{c)} When $40\%<r_S^{\rm{CER}} < 80\%$, both HESS and AESS can reduce LCOE, and more ambitious the carbon reduction target is, more obvious the impact on lowering LCOE.

\emph{d)} When $80\%<r_S^{\rm{CER}} < 100\%$, AESS has a more significant effect on reducing LCOE than HESS. Finally, under the goal of carbon neutrality, the contribution of HESS and AESS in reducing LCOE reaches $12.28\%$ and $14.59\%$, respectively.

In summary, carbon emission reduction targets are an intrinsic driver for developing AESS and HESS. The more compact the target is, the more HESS and AESS are required.





\section{Conclusions}
\label{sec:conclusions}
An MSEP model for decarbonizing the coal-fired supported renewable power system is proposed in this paper, as well as the corresponding DWD-CG solution method. A virtual internal trading-based LCOS assessment method is presented to well quantify LCOS for BESS, HESS, and AESS. A real-life UHVDC project from Gansu to Shandong in China is studied. The results indicate that BESS, HESS, and AESS play different roles in handling RES's stochasticity, volatility, and intermittence. Specifically, AESS has the ability and characteristics of monthly and seasonal regulation, known as seasonal energy storage. Furthermore, the proposed DWD-CG method is accurate and efficient for solving large-scale optimization problems with a controllable computation burden. 

Currently, although HESS and AESS can reduce the LCOE  to improve the economy of the system, their LCOSs are still far more than 1 RMB/kWh. How to design a reasonable electricity price incentive policy to realize the actual value of HESS and AESS, is a promising directions for future research.


\appendices
\section{LCOS Assessment of BESS, HESS, and AESS}
\label{sec:A1}
\setcounter{equation}{0}
\renewcommand{\theequation}{\ref{sec:A1}\arabic{equation}}
\setcounter{figure}{0}
\renewcommand{\thefigure}{\ref{sec:A1}\arabic{figure}}
\setcounter{table}{0}
\renewcommand{\thetable}{\ref{sec:A1}\arabic{table}}
First, the proposed system shown in Fig. \ref{fig:System_Topology} is divided into four parts: GEN part with facilities set $\Omega_{\rm{G}} = \left\{ \rm{W},\rm{S},\rm{CFPP}\right\}$, BESS part with facilities set $\Omega_{\rm{B}} = \left\{ B\right\}$, HESS part with facilities set $\Omega_{\rm{H}} = \left\{ \rm{AE},\rm{HS},\rm{FC}\right\}$, and AESS part with facilities set $\Omega_{\rm{A}} = \left\{ \rm{ASyn},\rm{ASto}\right\}$. GEN, BESS, HESS, and AESS sell electricity at prices $LCOE_{\rm{G}}$, $LCOS_{\rm{B}}$, $LCOS_{\rm{H}}$, and $LCOS_{\rm{A}}$, respectively. HESS sells hydrogen to AESS at price $LCOH$, shown in Fig. \ref{fig:LCOS_Assessment}. 

Then, the equation $NPV = 0$ for each part is presented as follows:
\begin{subequations}
  \begin{small}
  \begin{align}
    &NPV^{\rm{G}} = \left. \left(PVC^{\rm{inv}} + PVC^{\rm{O\&M}} - PVS \right) \right|_{\Omega_{\rm{G}}}  \nonumber\\
    &+ PVC^{\rm{reti}} + PVC^{\rm{coal}}- LCOE_{\rm{G}} *PVE^{\rm{G}} = 0 \label{eq:LCOS_1} \\
    &NPV^{\rm{B}} =\left. \left(PVC^{\rm{inv}} + PVC^{\rm{O\&M}} - PVS \right) \right|_{\Omega_{\rm{B}}}  \nonumber\\
    &+LCOE_{\rm{G}} *PVE^{\rm{B,ch}} - LCOS_{\rm{B}} *PVE^{\rm{B,disc}} = 0 \label{eq:LCOS_2} \\
    &NPV^{\rm{H}} =\left. \left(PVC^{\rm{inv}} + PVC^{\rm{O\&M}} - PVS \right) \right|_{\Omega_{\rm{H}}} \nonumber \\
    &+ PVC^{\rm{H,purch}} +LCOE_{\rm{G}} *PVE^{\rm{AE}} - PVR^{\rm{H,sell}}  \nonumber\\
    &- LCOH *PVE^{\rm{H,A}}  - LCOS_{\rm{H}} *PVE^{\rm{FC}} = 0 \label{eq:LCOS_3} \\
    &NPV^{\rm{A}} =\left. \left(PVC^{\rm{inv}} + PVC^{\rm{O\&M}} - PVS \right) \right|_{\Omega_{\rm{A}}} + PVC^{\rm{A,purch}}  \nonumber\\
    &+LCOE_{\rm{G}} *PVE^{\rm{AS}} + LCOH *PVE^{\rm{H,A}} \nonumber\\
    & - PVR^{\rm{H,sell}} - LCOS_{\rm{A}} *PVE^{\rm{AF}} = 0 \label{eq:LCOS_4} \\
    & LCOH = \kappa_{\rm{FC}} *LCOS_{\rm{H}} \label{eq:LCOS_5}
  \end{align}
  \end{small}
\end{subequations}
\noindent
\\where $PVE^{j} = \sum_{s\in \mathbb{S}} \sum_{t \in \mathbb{T}} \delta_{3}(s) P_{s,t}^{j}$ represents the present value of electricity/hydrogen over the whole planning horizon, and superscript $j$ appears in (\ref{eq:D_Power_Balance}), (\ref{eq:CFPP_1}), and (\ref{eq:D_HS_1}). In addition, $PVE^{\rm{G}}$ and $PVE^{\rm{G,D}}$ are denoted as follows:
\begin{small}
\begin{align}
  &PVE^{\rm{G}} = PVE^{\rm{W}} + PVE^{\rm{S}} + PVE^{\rm{CF}} - PVE^{\rm{curt}} \label{eq:LCOS_6} \\
  &PVE^{\rm{G,D}} = PVE^{\rm{G}} - PVE^{\rm{B,ch}} - PVE^{\rm{AE}} - PVE^{\rm{AS}}\label{eq:LCOS_7}
\end{align}
\end{small}

Finally, solving the system of linear equations (\ref{eq:LCOS_1})--(\ref{eq:LCOS_5}), we can obtain the corresponding solution of $LCOE_{\rm{G}}$, $LCOS_{\rm{B}}$, $LCOS_{\rm{H}}$, $LCOS_{\rm{A}}$, and $LCOH$. Furthermore, LCOE for UHVDC can also be calculated to assess the economics of the whole system, denoted as
\begin{small}
  \begin{align}
    &LCOE = \left(LCOE_{\rm{G}}*PVE^{\rm{G,D}} + LCOS_{\rm{B}}*PVE^{\rm{B,disc}} \right. \nonumber\\
    &\left. + LCOS_{\rm{H}}*PVE^{\rm{FC}} + LCOS_{\rm{A}}*PVE^{\rm{AF}} \right) / PVE^{\rm{UHVDC}} \label{eq:LCOS_8} 
  \end{align}
\end{small}

\begin{figure}[htb]
  \centering
  \includegraphics[width=3.46in]{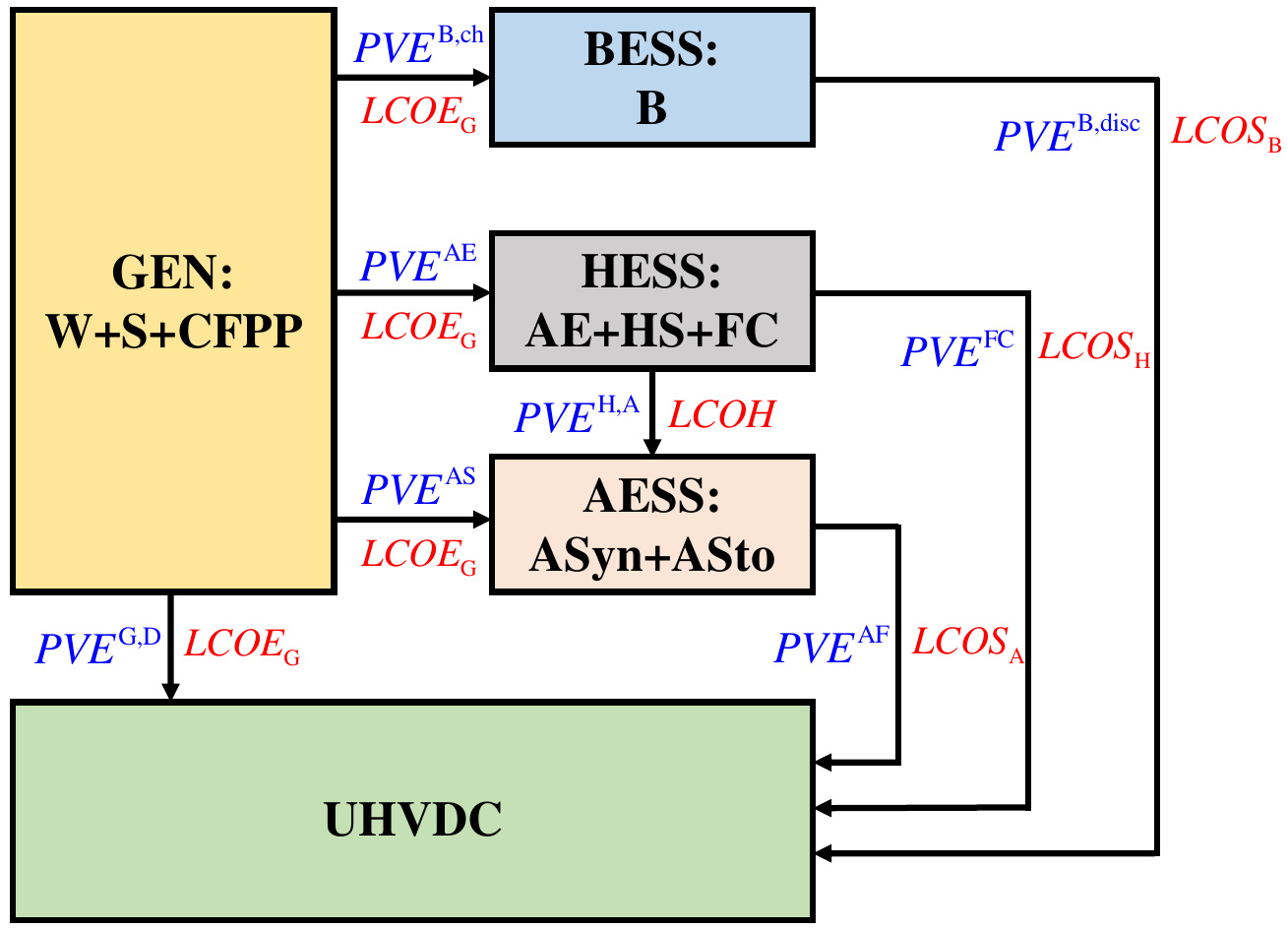}
  \caption{Virtual internal trading of electricity, hydrogen, and ammonia for LCOS assessment.} 
  \label{fig:LCOS_Assessment}
\end{figure}


\bibliographystyle{IEEEtran}
\bibliography{IEEEabrv,WSTES_MPEP_LDES}

\end{document}